\documentclass[11pt]{article}
\usepackage[usenames,dvipsnames,svgnames,table]{xcolor} % usenames : 16 basic colors,  dvipsnames : other 68 colors, svgnames : another 150 colors. It is used instead of \usepackage{color}. Can find descriptions for color in xcolor package. 

\usepackage{algorithm}
\usepackage{algorithmic}
\floatname{algorithm}{Algorithm}

\usepackage{longtable}
\usepackage{hhline}
\usepackage{float}
\newfloat{algorithm}{tb}{lop}

\usepackage{enumitem} % for \begin{description}
\usepackage{rotfloat} % For [H] option in sidewaystable
\usepackage{graphicx}
\usepackage{epsfig}
\usepackage{amssymb, amsmath, amsthm}
\usepackage{verbatim}
\usepackage{natbib}
\usepackage{authblk}
\usepackage{multirow}
\usepackage{subcaption} % for \begin{subtable}
\usepackage{array}
\newcolumntype{L}{>{\centering\arraybackslash}m{0.1\linewidth}}
\usepackage{hyperref}
\usepackage[usenames,dvipsnames,svgnames,table]{xcolor}
\usepackage{subcaption}
\usepackage{algorithm}
\usepackage{algorithmic}
\floatname{algorithm}{Algorithm}

\DeclareMathOperator*{\argmin}{argmin}

\newcommand{\calM}{\mathcal{M}}
\newcommand{\calN}{\mathcal{N}}
\newcommand{\calP}{\mathcal{P}}

\newcommand{\calW}{\mathcal{W}}

\newcommand{\bbP}{\mathbb{P}}
 
\newcommand{\bbR}{\mathbb{R}}

\newcommand{\bbE}{\mathbb{E}}
\newcommand{\bbN}{\mathbb{N}}

\newcommand{\bfx}{\mathbf{x}}

\newcommand{\Frechet}{Fr\'{e}chet }

\newcommand{\Rd}{\mathbb{R}^d}
\newcommand{\SPD}{\mathcal{N}_{0}^{d}}

\definecolor{wass-mean}{HTML}{CC79A7}
\definecolor{wass-meds}{HTML}{0072B2}
\definecolor{wass-class1}{HTML}{56B4E9}
\definecolor{wass-class2}{HTML}{D55E00}

\newtheorem{theorem}{Theorem}

\newtheorem{proposition}[theorem]{Proposition}

\begin{document}
	
\title{On the Wasserstein median of probability measures\footnote{This is an original manuscript of an article published by Taylor \& Francis in Journal of Computational and Graphical Statistics on August 29\textsuperscript{th}, 2024, available online: \url{https://doi.org/10.1080/10618600.2024.2374580}.
}\footnote{{\color{red} \textbf{Errata}}: Equation \eqref{eq:weiszfeld_map} is corrected, which used an incorrect form of the exponential map.}}
\author[1]{Kisung You}
\author[2]{Dennis Shung}
\author[2]{Mauro Giuffr\`{e}}
\affil[1]{Department of Mathematics, Baruch College}
\affil[2]{Department of Internal Medicine, Yale University School of Medicine}

\date{}

\maketitle
\begin{abstract}
The primary choice to summarize a finite collection of random objects is by using measures of central tendency, such as mean and median.
In the field of optimal transport, the Wasserstein barycenter corresponds to the \Frechet or geometric mean of a set of probability measures, which is defined as a minimizer of the sum of squared distances to each element in a given set with respect to the Wasserstein distance of order 2.
We introduce the Wasserstein median as a robust alternative to the Wasserstein barycenter. The Wasserstein median corresponds to the \Frechet median under the 2-Wasserstein metric.
The existence and consistency of the Wasserstein median are first established, along with its robustness property. 
In addition, we present a general computational pipeline that employs any recognized algorithms for the Wasserstein barycenter in an iterative fashion and demonstrate its convergence.
The utility of the Wasserstein median as a robust measure of central tendency is  demonstrated using real and simulated data. 
\end{abstract}

% ***************************************************************
\section{Introduction}\label{sec:intro}

The theory of optimal transport (OT) studies the mathematical structure for the space of probability measures and is one of the most prominent disciplines in modern data science. Originally introduced by Monge in the late 18th century \citep{monge_1781_MemoireTheorieDeblais}, the OT problem was revisited and generalized by Kantrovich, almost two centuries after its inception \citep{kantorovitch_1958_TranslocationMasses}. The framework of OT has long attracted much attention from theoretical perspectives \citep{ambrosio_2003_OptimalTransportationApplications, villani_2003_TopicsOptimalTransportation} and has gained broader recognition with the advent of efficient computational pipelines  \citep{cuturi_2013_SinkhornDistancesLightspeed, peyre_2019_ComputationalOptimalTransport}. The advancements in computational methods for OT have engendered much success in quantitative fields. In machine learning, for example, the OT framework has been applied to numerous tasks such as metric learning \citep{kolouri_2016_SlicedWassersteinKernels}, dimensionality reduction \citep{bigot_2018_CharacterizationBarycentersWasserstein},  domain adaptation \citep{courty_2017_OptimalTransportDomain, courty_2017_JointDistributionOptimal}, and improving the generative adversarial networks \citep{arjovsky_2017_WassersteinGenerativeAdversarial}, among others.

Statistics has also benefited from developments in OT by incorporating core machinery into various problems \citep{panaretos_2020_InvitationStatisticsWasserstein}. A few notable examples include parameter estimation via  minimum Wasserstein distance estimators \citep{bernton_2019_ParameterEstimationWasserstein}, sampling from the posterior without Markov chain Monte Carlo methods \citep{elmoselhy_2012_BayesianInferenceOptimal}, two-sample hypothesis testing in high dimensions \citep{ramdas_2017_WassersteinTwoSampleTesting}, approximate Bayesian computation \citep{bernton_2019_ApproximateBayesianComputation}, scalable Bayesian inference with a divide-and-conquer approach \citep{srivastava_2018_ScalableBayesBarycenter}, and many more. Along these methodological innovations, a large volume of theoretical research has been simultaneously conducted to establish foundational knowledge in topics such as rates of convergence for Wasserstein distances \citep{fournier_2015_RateConvergenceWasserstein} and optimal transport maps \citep{hutter_2021_MinimaxEstimationSmooth}, central limit theorems for the distance \citep{delbarrio_1999_CentralLimitTheorems, manole_2021_PluginEstimationSmooth}, and others. 

At this moment, we call for attention to one of the most fundamental quantities in statistics to which a large number of aforementioned methods are related -- the centroid.  Suppose we are given a set of real numbers $x_1, \ldots, x_n$ and their arithmetic mean $\bar{x} = \sum_{i=1}^n x_i/n$. The classical theory of statistics starts from examining how $\bar{x}$ behaves through the law of large numbers and the central limit theorem under certain conditions and proceeds to perform a number of inferential tasks thereafter. Not to mention distributional properties, $\bar{x}$ itself is often of importance to measure the central tendency for a given set of observations since it represents maximally compressed information for a random sample. It is well known that $\bar{x}$ is
heavily influenced by outliers. An alternative to the arithmetic mean   is the median, which is a minimizer of the sum of absolute distances to the data. The concepts of these centroids, initially studied in Euclidean spaces under robust statistics  \citep{huber_1981_RobustStatistics}, have been generalized to other contexts.

For instance, when data reside in a general metric space, these measures of central tendency correspond to the quantities called the \Frechet mean and \Frechet median, with their characteristics and properties well studied in the context of Riemannian manifolds \citep{kendall_1990_ProbabilityConvexityHarmonic, pennec_2006_IntrinsicStatisticsRiemannian, afsari_2011_RiemannianCenterMass, bhattacharya_2012_NonparametricInferenceManifolds}. In the field of OT, the concept of the  \Frechet mean is known as the Wasserstein barycenter, which minimizes the sum of squared 2-Wasserstein distances. Since the seminal work of \cite{agueh_2011_BarycentersWassersteinSpace}, its theoretical properties such as existence and uniqueness have been much studied along with computational studies that are of ongoing interests \citep{cuturi_2014_FastComputationWasserstein, dvurechenskii_2018_DecentralizeRandomizeFaster, claici_2018_StochasticWassersteinBarycenters, li_2020_ContinuousRegularizedWasserstein,xie_2020_FastProximalPoint, korotin_2021_ContinuousWasserstein2Barycenter}. Given this context, it is natural to inquire about the counterpart of the \Frechet median in OT.

This motivates our proposal and investigation of the Wasserstein median in response to the call. As its name entails, the Wasserstein median generalizes the \Frechet median onto the space of probability measures. A primary contribution of this paper is the formulation of a novel measure of central tendency in the field of OT, filling a gap in the literature. We prove its existence, consistency, and robustness. Another appealing contribution is that we present an algorithm for computing a Wasserstein median whose convergence is also proven. Our proposed algorithm is versatile, able to incorporate any existing algorithms for computing Wasserstein barycenters, effectively making it a meta-algorithm. Our numerical experiments provide ample evidence to highlight the importance of our proposed framework.

During the development of our study, we encountered the work of \cite{altschuler_2021_AveragingBuresWassersteinManifold}. This appears to be the first study recognizing the Wasserstein median as a direct extension of the geometric median to the space of probability measures. However, their scope is restricted to algorithmic aspects of the regularized Wasserstein median for a collection of Gaussian distributions on the Bures-Wasserstein manifold. Moreover, a number of the theoretical properties cited in the relevant literature are not directly applicable to the context due to the unfavorable characteristics of the Wasserstein space. Our distinctive contributions consist of formulating the concept in a more general context relevant to what most practitioners encounter on a daily basis, examining elementary theoretical properties, and proposing a generic class of algorithm beyond the Bures-Wasserstein manifold.

The rest of this paper is organized as follows. In Section \ref{sec:geometry}, we start our journey with a concise review of basic concepts in OT. We formulate the Wasserstein median problem and a generic algorithm in Section \ref{sec:main} along with relevant  theoretical results. In Section \ref{sec:special}, we discuss two special cases on how the Wasserstein median problem has its connection to the literature based on the arguments pertained to the computation. In Section \ref{sec:example}, we validate the proposed framework with simulated and real data examples. We conclude in Section \ref{sec:conclusion} with a discussion on issues and topics that help to pose potential directions for future studies.  All proofs and additional simulation results are provided in the Appendix. The algorithms proposed are implemented in the \textsf{R} package \texttt{T4transport}, available at \url{https://kisungyou.com/T4transport}. Codes to replicate a selection of examples are available at \url{https://github.com/kisungyou/papers}.

% ***************************************************************
\section{Background}\label{sec:geometry}

% 2-1. General W_p
We start this section by introducing basic definitions and properties of the Wasserstein space and the metric structure defined thereon. Let $\calP(\bbR^d)$ be the space of Borel probability measures on $\bbR^d$. The Wasserstein space of order $p$ on $\bbR^d$ is defined as 
\begin{equation*}
	\calP_p (\bbR^d) = \left\lbrace \mu \in \calP(\bbR^d) : \int_{\bbR^d} \|\bfx\|^p \, d\mu (\bfx) < \infty \right\rbrace,\quad p\geq 1,
\end{equation*}
where $\|\cdot \|$ is the standard norm in the Euclidean space. The distance for any $\mu, \nu \in \calP_p (\bbR^d)$ is defined as the minimum of total transportation cost 
\begin{equation}\label{eq:monge}
	\calW_p (\mu,\nu) = \left( \underset{T:\Rd \rightarrow \Rd}{\inf} \int_{\Rd \times \Rd } \|{\bf x}-T({\bf x})\|^p \, d\mu({\bf x})\right)^{1/p},
\end{equation}
for a measurable transport map $T:\Rd \rightarrow \Rd$ such that $T\# \mu = \nu$, i.e., for all Borel measurable sets $B$, $\nu(B) = \mu(T^{-1}(B))$. The equation \eqref{eq:monge} is known as the Monge formulation \citep{monge_1781_MemoireTheorieDeblais}. Although intuitive, the Monge formulation has some limitations in pragmatic settings in that it does not allow split of masses, which can be problematic when considering two discrete measures $\mu$ and $\nu$ of different cardinalities. Also, computation can be prohibitive because the optimization problem is considered for all mappings from $\Rd$ to $\Rd$ without any assumptions on $T$ except for the measurability. A relaxed version of the formulation was proposed by \cite{kantorovitch_1958_TranslocationMasses} as follows. In the Kantrovich formulation, the distance between two measures $\mu,\nu \in \calP_p (\Rd)$ is defined as
\begin{equation}\label{eq:kantrovich}
	\calW_p (\mu,\nu) = \left( \underset{\pi \in \Pi(\mu,\nu)}{\inf} \int_{\Rd \times \Rd } \|{\bf x}-{\bf y}\|^p  \, d\pi({\bf x},{\bf y})\right)^{1/p},
\end{equation}
where $\Pi(\mu,\nu)$ denotes the collection of all joint measures on $\Rd \times \Rd$ whose marginals are $\mu$ and $\nu$. Existence of an optimal joint measure from equation \eqref{eq:kantrovich} is guaranteed under mild conditions \citep{villani_2003_TopicsOptimalTransportation}.

% 2-2. Intuition

\begin{figure}[ht]
	\centering
	\begin{subfigure}[b]{0.32\textwidth}
		\centering
		\includegraphics[width=\textwidth]{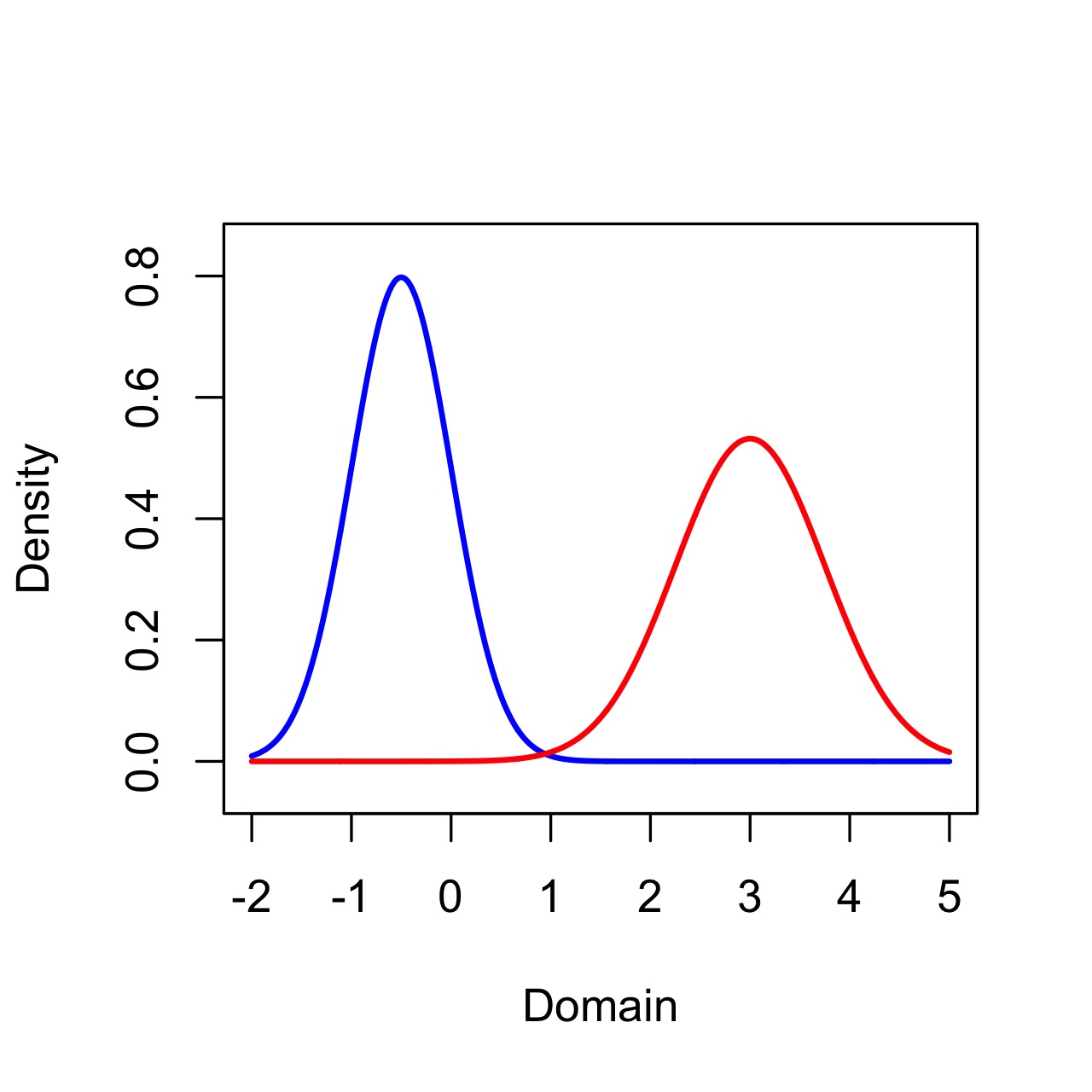}
		%\label{fig:y equals x}
	\end{subfigure}
	\begin{subfigure}[b]{0.32\textwidth}
		\centering
		\includegraphics[width=\textwidth]{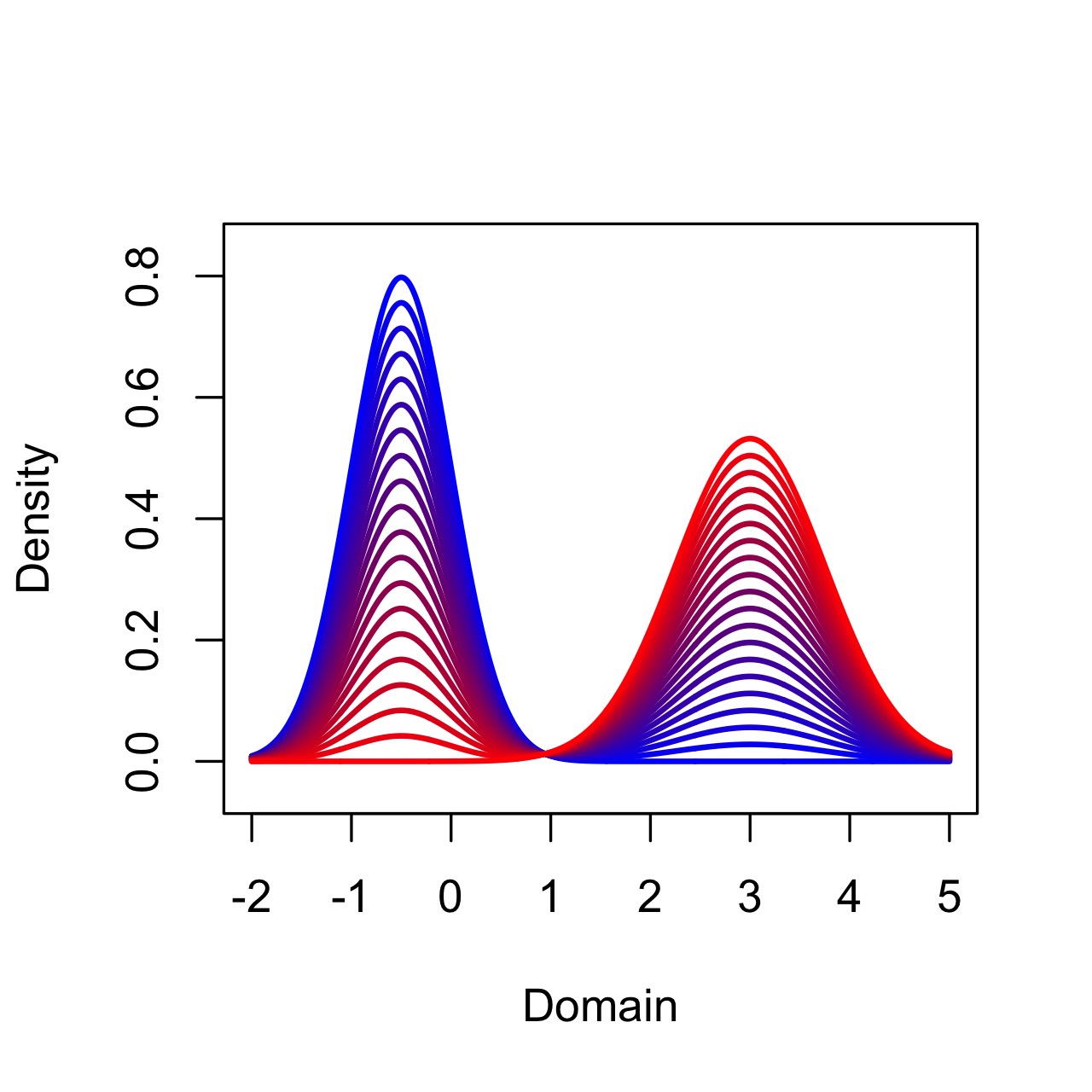}
		%\label{fig:three sin x}
	\end{subfigure}
	\begin{subfigure}[b]{0.32\textwidth}
		\centering
		\includegraphics[width=\textwidth]{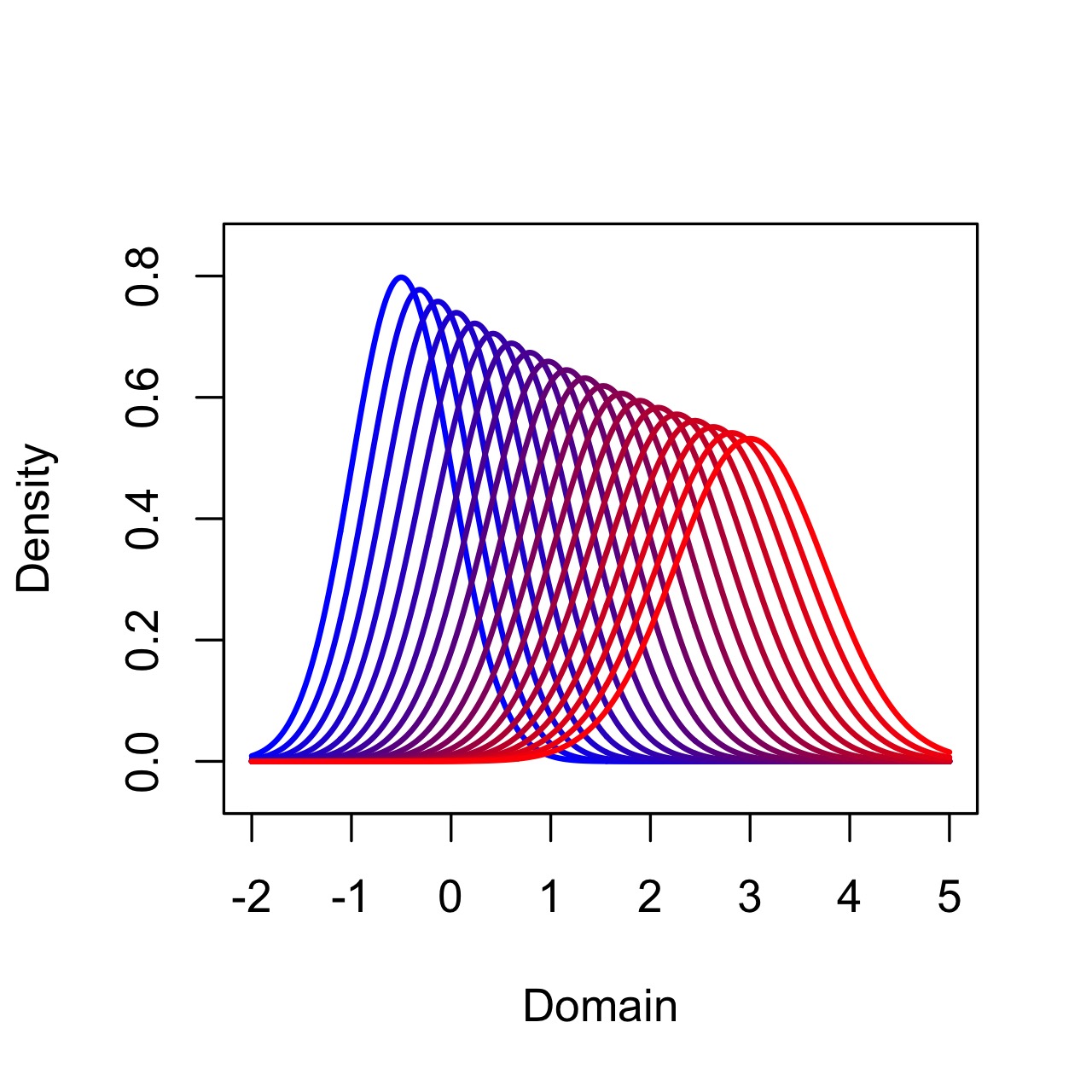}
		%\label{fig:three sin x}
	\end{subfigure}	
	\caption{A simple comparison of $L_2$ and OT geometries for two Gaussian distributions (left). Paths of interpolation are shown when $L_2$  (middle) and Wasserstein (right) geometries are considered.}
	\label{fig:simple_geometry}
\end{figure}

Before we delve into the formal introduction of geometric characterization, we aim to provide a visual comparison between the OT framework and functional perspectives within the space of probability measures. Consider a simple scenario involving two univariate Gaussian distributions that differ in their means and variances, as illustrated in Figure \ref{fig:simple_geometry}. A common method for interpolating between these two distributions is to take weighted averages of their density functions, given that these functions are elements of the $L_2$ space. Consequently, this method of interpolation results in a path characterized by vertical displacement. In contrast, the OT framework adopts a markedly different approach. True to its name, it focuses on the transportation of one measure to another. When we examine the densities of the interpolating distributions through the lens of OT, the transportation process appears to horizontally shift a measure while smoothly altering its overall shape.

% 2-3. Focus on W_2
We move our focus to mathematical facts on the Wasserstein distance at the minimal level. First, it is well known that  $\calW_p$ is not only a metric on $\calP_p (\bbR^d)$, but also metrizes the weak convergence of probability measures and the $p$-th moments \citep{villani_2003_TopicsOptimalTransportation}. Besides the metric structure, the space of measures $\calP_p(\Rd)$ has other desirable properties such as completeness and separability \citep{villani_2009_OptimalTransportOld}. Especially for the case of order $p=2$,  the 2-Wasserstein space $(\calP_2(\Rd), \calW_2)$ can be viewed as a complete Riemannian manifold of non-negative curvature \citep{otto_2001_GEOMETRYDISSIPATIVEEVOLUTION, ambrosio_2005_GradientFlowsMetric}.

For an absolutely continuous measure $\mu \in \calP_2(\Rd)$ and an arbitrary $\nu \in \calP_2(\Rd)$, $T_\mu^\nu$ stands for a  transport map from $\mu$ to $\nu$  and $I:\Rd \rightarrow \Rd$ for an identity map. Then, a curve $\mu_t = [I + t (T_\mu^\nu - I)]\#\mu $ for $t \in [0,1]$ that interpolates the two measures as $\mu_0 = \mu$ and $\mu_1 = \nu$ is known as McCann's interpolant \citep{mccann_1997_ConvexityPrincipleInteracting}, which is a constant-speed geodesic in $\calP_2 (\Rd)$. This perspective naturally leads to define the tangent space of $\calP_2(\Rd)$ at $\mu$ by
\begin{equation*}    
	\text{Tan}_\mu = \overline{
		\lbrace t (T_\mu^\nu - I) :~\nu \in \calP_2 (\Rd), ~t > 0\rbrace
	}^{L_2(\mu)}, 
\end{equation*}
which is a closed subset of the Banach space $L_2 (\mu) = \lbrace f:\Rd\rightarrow\Rd~\vert~\int_{\Rd} \|f({\bf x})\|^2 d\mu({\bf x}) < \infty\rbrace$. Using the map, one can define the  exponential map $\exp_\mu :\text{Tan}_\mu \rightarrow \calP_2(\Rd)$ and its inverse, called the logarithmic map $\log_\mu : \calP_2(\Rd) \rightarrow \text{Tan}_\mu$, as follows:
\begin{equation*}
	\exp_\mu (t(T_\mu^\nu - I)) = \lbrack t T_\mu^\nu + (1-t)I \rbrack\#\mu \quad \textrm{ and }\quad \log_\mu(\nu) = T_{\mu}^\nu - I.
\end{equation*}
In the rest of this paper, we restrict our attention to the $p=2$ case. For notational simplicity, we may interchangeably use "barycenter" and "median" to denote the Wasserstein barycenter and the Wasserstein median of order 2, provided this does not cause confusion in the context. We direct readers interested in the  characterization of the Wasserstein geometry to standard references such as \cite{villani_2003_TopicsOptimalTransportation} and \cite{ambrosio_2005_GradientFlowsMetric} for more details.

% ***************************************************************
\section{Methods}\label{sec:main}

\subsection{Problem Definition}
Let $\left\lbrace \mu_n \in \calP_2 (\bbR^d),~n=1,\ldots, N \right\rbrace $ be a collection of probability measures. Its Wasserstein median $\nu^*$ is defined as a minimizer of the functional
\begin{equation}\label{eq:problem_general}
	F(\nu) = \sum_{n=1}^N \pi_n \calW_2 (\nu, \mu_n),
\end{equation}
for nonnegative weights $(\pi_1,\ldots,\pi_N) \in \Delta_0^N := \lbrace {\bf x} \in \mathbb{R}^N~|~x_i>0,\, \sum_{i=1}^N x_i = 1\rbrace $. The minimization problem is well-defined since the Wasserstein distance $\calW_2$ is non-negative and continuous.  Given a sample version of the problem, it is natural to consider a population counterpart
\begin{equation}\label{eq:problem_general_population}
	F(\nu) = \bbE \calW_2 (\nu, \Lambda) = \int \calW_2 (\nu, \lambda) d\bbP(\lambda),
\end{equation}
with respect to a random measure $\Lambda$ on $\calP_2(\bbR^d)$ with distribution $\bbP$. We note that the functional \eqref{eq:problem_general} can be viewed as an expectation with respect to a discrete measure $\Lambda = \sum_{n=1}^N \pi_n \delta_{\mu_n}$ on the space of probability measures.

We present how the formulation of Wasserstein median is connected to the standard geometric median problem.  Suppose we are given two Dirac measures $\mu_{\bf x} = \delta_{\bf x}$ and $\mu_{\bf y} = \delta_{\bf y}$ on $\bbR^d$. By the definition \eqref{eq:kantrovich}, it is without doubt that  the Wasserstein distance of order 2 is reduced to the standard $L_2$ distance, i.e., $\mathcal{W}_2 (\mu_{\bf x},\mu_{\bf y}) = \|{\bf x}-{\bf y}\|$, since both measures are singletons. Therefore, if we further restrict the class of desired centroids to be Dirac measures, the Wasserstein median problem \eqref{eq:problem_general} translates to minimize the following objective function  
\begin{equation*}
	F(\nu) = \sum_{n=1}^N \pi_n \| {\bf x} - {\bf x}_n \|,
\end{equation*}
for some $\nu = \delta_{\bf x}$. This is indeed equivalent to finding a geometric median in the Euclidean setting. Hence, we may consider the proposed Wasserstein median as a direct generalization of geometric median to the space of probability measures.

\subsection{Computation}

We describe a geometric variant of the Weiszfeld algorithm \citep{weiszfeld_1937_PointPourLequel, fletcher_2009_GeometricMedianRiemannian} for the Wasserstein median problem by viewing the 2-Wasserstein space of probability measures $\calP_2 (\bbR^d)$ as a Riemannian manifold \citep{ambrosio_2021_LecturesOptimalTransport}. In this section, a sequence of minimizers is denoted as $v^{(t)}$ at an iteration $t$. 

We first describe a direct application of the geometric Weiszfeld algorithm on $ \calP_2(\bbR^d)$ using the transport-map based adaptation. First, the gradient of the cost function \eqref{eq:problem_general} can be written in terms of transport maps by
\begin{equation*}
	\nabla F(\nu) = -\sum_{n=1}^N \frac{\pi_n}{\calW_2 (\nu, \mu_n)} \log_v (\mu_n) = -\sum_{n=1}^N \frac{\pi_n}{\calW_2 (\nu, \mu_n)} \left\lbrack T_\nu^{\mu_n} - I \right\rbrack,
\end{equation*}
which resides on the tangent space of $\nu$. At every iteration, the geometric Weiszfeld algorithm updates a candidate by projecting the gradient back onto the Riemannian manifold $\calP_2 (\bbR^d)$ via an exponential map
\begin{equation*}
	v^{(t+1)} = \exp_{v^{(t)}} \left( -\tau^{(t)} \nabla F(\nu^{(t)}) \right),
\end{equation*}
with a varying step size $\tau^{(t)} = 1/\sum_{n=1}^N w_n^{(t)}$ where $w_n^{(t)} = \pi_n / \calW_2 (\nu^{(t)},\mu_n)$. Thus we may write the updating rule as 
\begin{equation}\label{eq:weiszfeld_map}
	\nu^{(t+1)} 
	= \exp_{v^{(t)}} \left(-\frac{1}{\sum_{n=1}^N w_n^{(t)}} \nabla F (\nu^{(t)}) \right)
	= \left(\sum_{n=1}^N  \tilde{w}_{n}^{(t)} T_{\nu^{(t)}}^{\mu_n}  \right) \# v^{(t)},
\end{equation}
for a normalized weight vector $\tilde{w}_{n}^{(t)} = w_n^{(t)}/\sum_{n=1}^N w_n^{(t)}$. 
The geometric Weiszfeld algorithm follows a descent path by varying amounts of step size, which is determined by a fixed rule at each iteration. This mechanism helps to reduce the total computational complexity since a standard line search method requires repeated evaluations of the functional.

The transport map-based Weiszfeld algorithm, however, may not be a practical choice from a computational point of view. The algorithm requires to compute $N$ transport maps $T_{v^{(t)}}^{\mu_n}$ at every iteration, which is an expensive procedure in general and an area of active research by itself. When the measures are all discrete, this may attenuate overall computational burden although the naive implementation of the map-based algorithm is expected to continuously update an iterate whose cardinality is equal to the sum of cardinalities of all measures under consideration. 

Instead, we take an alternative view of the Weiszfeld algorithm as an iteratively reweighted least squares (IRLS) method \citep{lawson_1961_ContributionsTheoryLinear, rice_1964_ApproximationsFunctions, osborne_1985_FiniteAlgorithmsOptimization}. The IRLS algorithm is a type of gradient descent methods that solves an optimization problem involving a $p$-norm. A central idea behind the IRLS is to solve an abstruse problem by iteratively solving a sequence of weighted least square problems. Its simplicity has allowed widespread adoption of the algorithm in a number of applications such as  compressed sensing \citep{gorodnitsky_1997_SparseSignalReconstruction, chartrand_2008_IterativelyReweightedAlgorithms, daubechies_2010_IterativelyReweightedLeast} and parameter estimation in generalized linear models \citep{nelder_1972_GeneralizedLinearModels, mccullagh_1998_GeneralizedLinearModels}.

In the Wasserstein median problem, we adapt the IRLS algorithm as follows. First, consider the following minimization functional at iteration $t$,
\begin{equation}\label{eq:IRLS_eachstep}
	G_t(\nu) =  \frac{1}{2}\sum_{n=1}^N \frac{\pi_n}{\calW_2 (\nu^{(t)}, \mu_n)} \calW_2^2 (\nu, \mu_n) = \frac{1}{2} \sum_{n=1}^N {w_n^{(t)}} \calW_2^2 (\nu, \mu_n).
\end{equation}
It is immediate to observe  that $G_t (\nu^{(t)})$ is half the minimum value attained at the $t$-th iteration. Using similar machineries, the gradient of $G_t (\nu)$ is written as 
\begin{equation*}
	\nabla G_t (\nu) = - \sum_{n=1}^N w_n^{(t)} \log_v (\mu_n).
\end{equation*}
One can directly check from the first-order condition that the updating rule \eqref{eq:weiszfeld_map} of the Weiszfeld algorithm is equal to solving the minimization problem \eqref{eq:IRLS_eachstep} iteratively. Furthermore, scaling every weight $w_n^{(t)},~n=1,\ldots,N$ by any positive scalar does not affect the solution of \eqref{eq:IRLS_eachstep}. Hence, we can rephrase the IRLS problem at iteration $t$,
\begin{equation}\label{eq:IRLS_subproblem}
	\min_{\nu}	G_t (\nu) = \min_{\nu} \sum_{n=1}^N \tilde{w}_n^{(t)} \calW_2^2 (\nu, \mu_n),
\end{equation}
which corresponds to the standard formulation of the Wasserstein barycenter problem. This implies that the Wasserstein median problem is equivalent to solving a sequence of Wasserstein barycenter problems. The described pipeline is summarized in Algorithm \ref{code:general_irls}. We make a remark on the computational complexity of the Algorithm \ref{code:general_irls}. If a solver for the Wasserstein barycenter problem has complexity $O(M)$, then the Wasserstein median computation has complexity of $O(KM)$ where $K$ is the number of iterations til convergence, which indicates its inherent bottleneck. This implies that performance of the proposed algorithm heavily depends on efficiency of a barycenter solver and intrinsic geometry of data space that governs the number of iterations.

\begin{algorithm}[t]
	\caption{Wasserstein median computation by IRLS.}
	\label{code:general_irls}
	\begin{algorithmic}
		\REQUIRE a collection of probability measures $\lbrace \mu_n \rbrace_{n=1}^N$, weights $(\pi_1,\ldots,\pi_N)\in\Delta_0^N$.
		\ENSURE $\nu^* = \argmin_{\nu \in \calP_2 (\bbR^d)} \sum_{n=1}^N \pi_n \calW_2 (\nu, \mu_n)$.
		\STATE Initialize $\nu^{(0)} \in \calP_2 (\bbR^d)$.
		\REPEAT 
		\STATE $w_n^{(t)} = \pi_n / \calW_2 (\nu^{(t)}, \mu_n)$.
		\STATE $\tilde{w}_n^{(t)} = w_n^{(t)}/\sum_{n=1}^N w_n^{(t)}$.
		\STATE $\nu^{(t+1)} = \argmin_\nu G_t (\nu) = \argmin_\nu \sum_{n=1}^N \tilde{w}_n^{(t)} \calW_2^2 (\nu, \mu_n)$ for $\nu \in \calP_2 (\bbR^d)$.
		\IF{
			$\nu^{(t+1)}$ is equal to one of $\mu_n$'s or $\nu^{(t+1)}$ is close to $\nu^{(t)}$
		}
		\STATE return $\nu^{(t+1)}$ and stop.
		\ENDIF 
		\UNTIL convergence.
	\end{algorithmic}
\end{algorithm}

We close this section by elaborating practical components in our computational pipeline. First is the choice of a starting point $\nu^{(0)}$. It seems reasonable to choose the Wasserstein barycenter as an initial starting point $\nu^{(0)}$ since both barycenter and median are centroids and they should be sufficiently close to each other if not much degree of outlying observations is expected. The barycenter computation, however, is already an expensive operation that should be run iteratively under our framework. As a remedy, we opted to select  one of the given probability measures that minimizes the sum of the distances as $\nu^{(0)}$. This is motivated from the fact that the first step at each iteration  involves attaining $w_n^{(t)} = \pi_n / \calW_2 (\nu^{(t)}, \mu_n)$, which accompanies distance computation between an iterate $\nu^{(t)}$ and all given measures $\mu_n$. Hence, our strategy removes the need for evalution of all cross distances in the first iteration. Under this strategy, one needs to make slight adjustment to the algorithm. Suppose $\mu_k$ for some $1\leq k\leq N$ was selected as $\nu^{(0)}$, then $w_k^{(0)}$ is not a finite number because of the division by 0. This necessitates the removal of $\mu_k$ from the computation of the next iterate, while re-normalizing $\tilde{w}_n^{(0)}$ to ensure no weight becomes singular.

The second component pertains to the  stopping conditions in Algorithm \ref{code:general_irls}. In the literature of Weiszfeld algorithm, the iterative process is terminated when an iterate $\nu^{(t+1)}$ corresponds to one of the inputs. This is because an adjusted weight in the following iteration diverges as a denominator becomes zero. An ad hoc remedy would be padding denominators of zeros with a very small number, which is equivalent to assigning an exceptionally large weight on the coinciding observation. In our experiments, we observed that such an adjusting mechanism may not result in a meaningful update of an iterate in many cases. Hence, we introduced another stopping criterion when two consecutive iterates are
`close', which may indicate multiple options. For instance, the most straightforward rule to terminate by proximity of iterates would be in terms of their Wasserstein distance, i.e., when $\calW_2 (\nu^{(t)}, \nu^{(t+1)}) < \epsilon$ for some small $\epsilon > 0$. This is justified from the fact that $\calP_2 (\bbR^d)$ is a complete metric space and convergence is upheld when a sequence is Cauchy. This stopping rule, however, may not be practical since computation of Wasserstein distance is a relatively expensive operation. A simpler alternative is to use a stopping rule based on the distances in the ambient space \citep{you_2022_ParameterEstimationModelbased}. This can be justified by the fact that for any two points $x,y$ on a complete Riemannian manifold $\calM$ embedded in some Euclidean space, $\lim_{x \rightarrow y} d_\calM(x,y)/\|x-y\| = 1$ where $d_\calM(\cdot, \cdot)$ is the geodesic distance on $\calM$. For instance, we used the Frobenius distance between two consecutive covariance matrices for centered Gaussian measures experiment and the standard $L_2$ distance between marginal vectors in all other examples. We found no meaningful differences between solutions when stopping criteria were set by these strategies under the Cauchy argument.

\subsection{Theory}
With the above formulation and a generic algorithm, we present theoretical properties of the Wasserstein median problem. For the well-defined statistic, the first is to show  the existence of a solution to the problem \eqref{eq:problem_general_population}. Our strategy follows an argument of \cite{bigot_2018_CharacterizationBarycentersWasserstein} that showed existence of a Wasserstein barycenter.
\begin{proposition}\label{theory:median_existence}
	The functional $F:\calP(\bbR^d)\rightarrow\bbR$ such that $F(\nu) = \bbE \calW_2 (\nu, \Lambda)$ with respect to some random measure $\Lambda$ on $\calP_2(\Rd)$ admits a minimizer $\nu^*$.	
\end{proposition}

The use of a random measure $\Lambda$ on $\calP_2 (\bbR^d)$ with distribution $\bbP$ is supported by a fact that the space itself is a complete and separable metric space  \citep{villani_2003_TopicsOptimalTransportation}, which allows to define the ``second degree'' Wasserstein space $\calP_2(\calP_2 (\bbR^d))$. This perspective motivated investigation of consistency for the Wasserstein barycenter \citep{legouic_2017_ExistenceConsistencyWasserstein} as follows. Consider a sequence of measures on $\calP_2(\bbR^d)$, all of which admit barycenters. If the sequence converges to some measure, a natural question is whether the sequence of corresponding barycenters also converges to that of a limiting measure, an answer of which turned out to be positive. Our claim that this phenomenon is also observed for the case of the Wasserstein median as follows.

% \textbf{this is Theorem 3 of \cite{legouic_2017_ExistenceConsistencyWasserstein}}.
\begin{proposition}\label{theory:median_consistency}
	Let $\lbrace \Lambda_j \rbrace$ be a sequence of probability measures on $\calP_2 (\bbR^d)$ that admits a corresponding set of Wasserstein medians $\lbrace \nu_j\rbrace $ for all $j\in \bbN$. Suppose $\calW_2 (\Lambda_j, \Lambda) \rightarrow 0$ for some $\Lambda \in \calP_2 (\calP_2 (\bbR^d))$ as $j\rightarrow \infty$. Then the sequence $\lbrace \nu_j \rbrace$ has a limit point $\nu$, which is a Wasserstein median of $\Lambda$. 
\end{proposition}

% the first corollary is from  \citep{legouic_2017_ExistenceConsistencyWasserstein}.
Consistency of a Wasserstein median implies immediate corollaries which we delineate in an informal manner. First, assume that $\Lambda$ admits a unique median and  $\mathbb{E}\calW_2(\nu, \Lambda_n) < \infty$ for all $n$.  If a sequence of measures $\lbrace \Lambda_j  \rbrace$ that converges to $\Lambda$, the sequence of Wasserstein medians $\lbrace \nu_j \rbrace$ also converges to the unique median of $\Lambda$.  This further implies that any convergent algorithm in the sense of Cauchy can successfully estimate the  Wasserstein median if uniqueness conditions are ensured. Another implication is the law of large numbers for the Wasserstein median. Consider $\Lambda_j$ as an empirical version of $\Lambda$ of increasing cardinality. Proposition 2.2.6 of \cite{panaretos_2020_InvitationStatisticsWasserstein} claims that $\Lambda_n$ converges to $\Lambda$ almost surely under the 2-Wasserstein metric if and only if $\Lambda \in \calP_2 (\calP_2 (\bbR^d))$. Application of Proposition \ref{theory:median_consistency} on top of almost sure convergence of $\Lambda_j \rightarrow \Lambda$ gives a rise to the law of large numbers in a weak sense.

An advantage of the Wasserstein median over the barycenter is robustness to outliers, which is an overarching property for a class of medians defined across different sample spaces. A popular measure of robustness for centroid estimators is the breakdown point \citep{hampel_1971_GeneralQualitativeDefinition}. The breakdown point measures the proportion of largely deviated observations within the data that an estimator can handle without yielding an incorrect result; the higher the breakdown point, the more robust the estimator \citep{rousseeuw_1987_RobustRegressionOutlier}. Let $T$ be a centroid estimator and $X = \lbrace x_1, \ldots, x_n \rbrace$ a random sample in some metric space $(M, d)$. The breakdown point is defined as 
\begin{equation*}
	\epsilon^* (T, X) = \frac{1}{n} \min \left\lbrace  k \in \lbrace 1, \ldots, n\rbrace: \sup_{Y_{n,k}} d(T(X), T(Y_{n,k})) = \infty \right\rbrace,
\end{equation*}
where $Y_{n,k}$ is the empirical distribution of a replacement sample with at least $n-k$ points from the original sample $X$. That is, $Y_{n,k}$ contains at least $n-k$ observations drawn without replacement from $X$ and the rest consists of arbitrary points from $M$. The finite-sample breakdown point for an unbounded Riemannian manifold was examined in \cite{fletcher_2009_GeometricMedianRiemannian}. This result allows direct extension to the 2-Wasserstein space, which we introduce for completeness.

\begin{theorem}[Theorem 2 of \cite{fletcher_2009_GeometricMedianRiemannian}]\label{theory:BreakdownPoint}
	Let $\mathcal{X} = \lbrace \mu_1, \ldots, \mu_N \rbrace$ be a collection of probability measures with finite second moments that is contained in a convex subset $\mathcal{U}$ of $\calP_2(\mathbb{R}^d)$ with $\textrm{diam}(\mathcal{U}) = \infty$. Then the Wasserstein median  $T_{med}$ has a breakdown point $\epsilon^*(T_{med}, \mathcal{X}) = \lfloor (N+1)/2\rfloor/N$.
\end{theorem}

Now, we turn our attention to the proposed algorithm for the Wasserstein median. For an iterative algorithm, it is of utmost importance to show convergence of the prescribed procedure to a set of minima by arguing that the updating procedure induces a decreasing sequence in $\calP_2 (\bbR^d)$ with respect to the functional. The following theorem establishes convergence of the proposed algorithm to a set of local minima. 

\begin{theorem}\label{thm:IRLS_convergence}
	Let $\lbrace \mu_n\rbrace_{n=1}^N \subset \calP_2 (\bbR^d)$ be a collection of probability measures that are absolutely continuous with respect to the Lebesgue measure and  weights $(\pi_1,\ldots,\pi_N) \in \Delta_0^N$. Define an updating function $U:\calP_2(\mathbb{R}^d) \rightarrow \calP_2(\mathbb{R}^d)$, which is obtained by minimizing $G_t (\nu)$ as in Equation \eqref{eq:IRLS_eachstep}. 	For an initial measure $\nu^{(0)}\in \calP_2 (\bbR^d)$ such that $F(\nu^{(0)}) < \infty$, the sequence $\lbrace \nu^{(t)}\rbrace$ for $t\in\bbN$  induced by the Algorithm \ref{code:general_irls} converges to a stationary point in a set $S = \lbrace \nu\in\calP_2(\bbR^d)~|~ F(U(\nu)) = F(\nu)\rbrace$.
\end{theorem}

The last theoretical contribution is on the uniqueness of an attained Wasserstein median. We previously presented the existence and consistency of the Wasserstein median with respect to a population version of its formulation; however, we did not establish uniqueness due to the unfavorable topological properties of the Wasserstein space. Yet, determining this theoretical property in a more realistic setting, where a collection of objects has finite cardinality, is of great interest. We introduce a consequence of Theorem \ref{thm:IRLS_convergence} that guarantees the uniqueness of the solution under mild conditions.

\begin{proposition}\label{thm:FiniteUniqueness}
	Let $\lbrace \mu_n \rbrace_{n=1}^N \subset \calP_2(\bbR^d)$ be a finite collection of probability measures that admits a unique Wasserstein barycenter, i.e., the functional $\sum_{n=1}^N \calW_2^2 (\mu_n, \nu)$ has a unique minimizer. Then, the Wasserstein median attained by Algorithm \ref{code:general_irls} is unique. 
\end{proposition}

We close this section with a remark: the uniqueness of the Wasserstein barycenter, as established by \cite{agueh_2011_BarycentersWassersteinSpace, legouic_2017_ExistenceConsistencyWasserstein}, is guaranteed provided that at least one of the $\mu_n$ measures vanishes on small sets. This naturally follows from Theorem \ref{thm:IRLS_convergence}, as any absolutely continuous probability measure, by definition, vanishes on sets of Lebesgue measure zero.

% ***************************************************************
\section{Special Cases}\label{sec:special}

The proposed framework of Wasserstein median applies to a general class of probability measures in $\calP_2(\mathbb{R}^d)$. In this section, we consider two specific classes of one-dimensional distributions and  Gaussian measures. These have their own favorable properties in common over arbitrary elements of $\calP_2(\bbR^d)$ that have made them as appealing objects with which to work.

\subsection{One-dimensional Distributions}

The first special case is $d=1$. Let $\mu_1$ and $\mu_2$ be two probability measures on $\bbR$ with cumulative distribution functions $F_1$ and $F_2$. The 2-Wasserstein distance between two measures is defined as
\begin{equation*}
	\calW_2 (\mu_1, \mu_2) = \left( \int_{0}^1 |F_1^{-1}(x) - F_2^{-1}(x)|^2 dx \right)^{1/2} = \|F_1^{-1} - F_2^{-1}\|_{L_2(0,1)} = \|Q_1 - Q_2 \|_{L_2(0,1)},
\end{equation*}
where the $Q_i$'s are quantile or inverse cumulative distribution functions, i.e., $Q_i (z) = \inf \lbrace x \in \bbR~:~ F_i(x) \geq z\rbrace$. Given a collection of one-dimensional probability measures $\lbrace \mu_n \rbrace_{n=1}^N$ with quantile functions $\lbrace Q_n\rbrace_{n=1}^N$, the Wasserstein median problem is translated into finding a \Frechet median in the function space  $L_2 (0,1)$ as follows, by omitting the subscript for notational simplicity throughout this section,
\begin{equation*}
	Q^* = \argmin_{Q} \sum_{n=1}^N \pi_n \|Q - Q_n \|.
\end{equation*}
Since any Borel probability measure on $\bbR$ can be uniquely determined by its cumulative distribution function, the minimizer $Q^*$ fully characterizes the Wasserstein median $\nu^*$ of the given probability measures. Indeed, it is a well-known fact that $\calP_2 (\bbR^d)$ has non-negative curvature in the sense of Alexandrov while $d=1$ endows the space with zero curvature. 

In the IRLS formulation, updates are obtained by a sequence of minimization problems. At iteration $t$, the optimization problem is written as 
\begin{equation}\label{problem:IRLS_functional}
	\min_Q G_t (Q)  = \min_Q \sum_{n=1}^N \tilde{w}_n^{(t)} \|Q - Q_n\|^2, 
\end{equation}
where $w_n^{(t)} = \pi_n/\|Q^{(t)} - Q_n\|$ and $\tilde{w}_n^{(t)} = w_n^{(t)}/\sum_{n=1}^N w_n^{(t)}$. The solution of \eqref{problem:IRLS_functional} is explicitly expressed as a weighted sum of $Q_n$'s,
\begin{equation*}
	Q^{(t+1)} = \argmin_Q G_t(Q) = \sum_{n=1}^N \tilde{w}_n^{(t)} Q_n,
\end{equation*}
which can be easily shown by solving for the Gateaux derivative of $G_t (Q)$. This accounts for updating the coefficients only without actually computing the quantity of interest at every iteration. We note that the Wasserstein median problem for one-dimensional probability measures admits a unique minimizer if the quantile functions $Q_1,\ldots,Q_N$ are not collinear \citep{vardi_2000_MultivariateL1medianAssociated, minsker_2015_GeometricMedianRobust}.

\subsection{Gaussian Distributions}

Another case of special interest is the space of non-degenerate Gaussian measures $\calN_2(\Rd)$, which is a strict subset of $\calP_2(\Rd)$. Given two multivariate Gaussian measures $\mu_1 = N({\bf m}_1,{\bf \Sigma}_1)$ and $\mu_2 = N({\bf m}_2, {\bf \Sigma}_2)$, the 2-Wasserstein distance is defined by
\begin{equation*}
	\calW_2 (\mu_1, \mu_2) = \left\lbrace  \| {\bf m}_1 - {\bf m}_2 \|^2 + \text{Tr} \left( {\bf \Sigma}_1 + {\bf \Sigma}_2 - 2 ({\bf \Sigma}_1{\bf \Sigma}_2)^{1/2}\right)\right\rbrace^{1/2},
\end{equation*}
which has appeared  in several studies \citep{dowson_1982_FrechetDistanceMultivariate, olkin_1982_DistanceTwoRandom, givens_1984_ClassWassersteinMetrics, knott_1984_OptimalMappingDistributions}. \cite{mccann_1997_ConvexityPrincipleInteracting} showed that the space of Gaussian measures is a totally geodesic submanifold of $\calP_2(\Rd)$. Based on the observation of $\calN_2(\Rd)$ as a submanifold, a Riemannian metric was proposed whose induced geodesic distance coincides with the 2-Wasserstein distance  \citep{takatsu_2011_WassersteinGeometryGaussian} and corresponding geometric operations were subsequently studied \citep{malago_2018_WassersteinRiemannianGeometry}. A Gaussian measure $N({\bf m}, {\bf \Sigma})$ is parametrized by two finite-dimensional parameters,  a mean vector ${\bf m}$ and a  covariance matrix ${\bf \Sigma}$. It is straightforward to see that  $\calN_2(\Rd)$ is a product metric space of the Euclidean space $\Rd$ with trivial geometry and the space of centered Gaussian measures $\calN_0^d$ endowed with a Riemannian metric. Hence, we limit our treatment of Gaussian distributions to $\SPD$ and interchangeably denote a centered Gaussian measure $N({\bf 0},{\bf \Sigma})$ as $N({\bf \Sigma})$.

Let $\mu_n = N({\bf \Sigma}_n)$ for $n=1,\ldots,N$ be a collection of non-degenerate centered Gaussian measures on $\Rd$. The Wasserstein median $\nu^* = N({\bf \Sigma}^*)$ is a minimizer of the equivalent functional
\begin{equation}\label{eq:gaussian_functional_abstract}
	F({\bf \Sigma}) = \sum_{n=1}^N \pi_n W_2 ({\bf \Sigma}, {\bf \Sigma}_n) = \sum_{n=1}^N \pi_n \left\lbrace   \text{Tr} \left( {\bf \Sigma} + {\bf \Sigma}_n - 2 ({\bf \Sigma}{\bf \Sigma}_n)^{1/2}\right)\right\rbrace^{1/2},
\end{equation}
for a distance function $W_2$ restricted on $\calN_0^d\times \calN_0^d$,
\begin{equation*}
	W_2^2({\bf \Sigma}_1, {\bf \Sigma}_2) := \calW_2^2(N({\bf 0},{\bf \Sigma}_1), N({\bf 0},{\bf \Sigma}_2)) = \text{Tr}({\bf \Sigma}_1+{\bf \Sigma}_2-2({\bf \Sigma}_1{\bf \Sigma}_2)^{1/2}).
\end{equation*}
At iteration $t$, a subproblem from the iterative IRLS formulation of minimizing \eqref{eq:gaussian_functional_abstract} is written as 
\begin{equation}\label{eq:IRLS_subproblem_Gaussian}
	\min_{{\bf \Sigma}} G_t ({\bf \Sigma}) = \min_{{\bf \Sigma}} \sum_{n=1}^N \tilde{w}_n^{(t)} W_2^2 ({\bf \Sigma}, {\bf \Sigma}_n),
\end{equation}
for $w_n^{(t)} = \pi_n / W_2({\bf \Sigma}^{(t)}, {\bf \Sigma}_n)$ and $\tilde{w}_n^{(t)} = w_n^{(t)}/\sum_{n=1}^N w_n^{(t)}$. We notice that the subproblem \eqref{eq:IRLS_subproblem_Gaussian} corresponds to computing the Wasserstein barycenter on $\calN_0^d$. Here, we introduce two variants of fixed-point approaches to solve the problem. In order to minimize confusion induced by notation, we consider the following Wasserstein barycenter problem,
\begin{equation*}
	\underset{{\bf S} \in \calN_0^d}{\min}\, \sum_{n=1}^N \lambda_n W_2^2 ({\bf S}, {\bf \Sigma}_n),
\end{equation*}
for $(\lambda_1,\ldots,\lambda_N) \in \Delta_0^N$ and use the subscript ${\bf S}_{(k)}$ to denote iterations in the barycenter problem. Given an appropriate non-singular starting point ${\bf S}_{(0)} \in \calN_0^d$, the first algorithm by \cite{ruschendorf_2002_NCouplingProblem} updates an iterate by
\begin{equation}\label{eq:barycenter_ruschendorf}
	{\bf S}_{(k+1)} = \sum_{n=1}^N \lambda_n \left({\bf S}_{(k)}^{1/2} \, {\bf \Sigma}_n \, {\bf S}_{(k)}^{1/2}\right)^{1/2},
\end{equation}
where ${\bf S}^{1/2}$ is positive square root of a symmetric, positive-definite matrix. Another iterative algorithm was proposed by \cite{alvarez-esteban_2016_FixedpointApproachBarycenters} where a single-step update is performed via 
\begin{equation}\label{eq:barycenter_alvarez}
	{\bf S}_{(k+1)} = {\bf S}_{(k)}^{-1/2}  \left(\sum_{n=1}^N \lambda_n  \left({\bf S}_{(k)}^{1/2}\, {\bf \Sigma}_n \,{\bf S}_{(k)}^{1/2}\right)^{1/2}\right)^2 {\bf S}_{(k)}^{-1/2}.
\end{equation}
While the latter update rule of \eqref{eq:barycenter_alvarez} seems  more complicated and incurs larger computational costs than \eqref{eq:barycenter_ruschendorf}, it was shown that a limit point of the iteration is a consistent estimator of the true barycenter \citep[Theorem 4.2]{alvarez-esteban_2016_FixedpointApproachBarycenters}. A computational routine for Wasserstein median of centered Gaussian measures under the IRLS formulation is summarized in Algorithm \ref{code:general_gaussian}.

\begin{algorithm}[h]
	\caption{Wasserstein median of centered Gaussian measures.}
	\label{code:general_gaussian}
	\begin{algorithmic}
		\REQUIRE a collection of full-rank covariance matrices $\lbrace \Sigma_n \rbrace_{n=1}^N$, weights $(\pi_1,\ldots,\pi_N)\in\Delta_0^N$.
		\ENSURE $\Sigma^* = \argmin_{\Sigma \in \calN_0^d} \sum_{n=1}^N \pi_n W_2 (\Sigma, \Sigma_n)$.
		\STATE Initialize $\Sigma^{(0)}$.
		\REPEAT 
		\STATE $w_n^{(t)} = \pi_n / W_2 (\Sigma^{(t)}, \Sigma_n)$.
		\STATE $\tilde{w}_n^{(t)} = w_n^{(t)}/\sum_{n=1}^N w_n^{(t)}$.
		\STATE Set $S_{(0)} = \Sigma^{(t)}$.
		\REPEAT
		\STATE Update an iterate $S_{(k)}$ by 
		\begin{equation*}
			S_{(k+1)}= \begin{cases}
				\sum_{n=1}^N \tilde{w}_n^{(t)}  \left(S_{(k)}^{1/2} \, \Sigma_n \, S_{(k)}^{1/2}\right)^{1/2},\,\, \textrm{ or } \\[10pt]
				S_{(k)}^{-1/2}  \left(\sum_{n=1}^N \tilde{w}_n^{(t)}  \left(S_{(k)}^{1/2}\, \Sigma_n \,S_{(k)}^{1/2}\right)^{1/2}\right)^2 S_{(k)}^{-1/2}
			\end{cases}
		\end{equation*}
		\UNTIL convergence and denote the limit point as $S_{(*)}$.
		\STATE Update $\Sigma^{(t+1)} = S_{(*)}$.
		\IF{
			$\Sigma^{(t+1)}$ is equal to one of $\Sigma_n$'s
		}
		\STATE return $\Sigma^{(t+1)}$ and stop. 
		\ENDIF 
		\UNTIL convergence.
	\end{algorithmic}
\end{algorithm}

% ***************************************************************
\section{Examples}\label{sec:example}

In this section, we demonstrate the effectiveness of the Wasserstein median as a robust measure of central tendency using simulated and real data examples. For the simulation study, a comparison is made between the Wasserstein median and the barycenter when the finite random sample contains  outliers. We extend the comparison to three real-data examples for parameter estimation, classification, and cluster analysis on different types of measure-valued data. 

\subsection{Simulated Example}

\begin{figure}[t]
	\centering
	\begin{subfigure}[b]{0.40\textwidth}
		\centering
		\includegraphics[width=\textwidth]{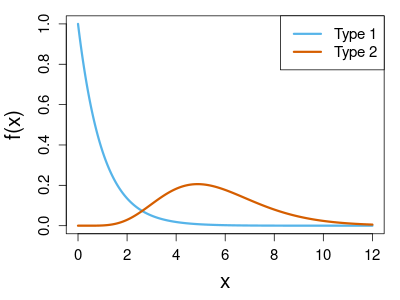}
		%\label{fig:y equals x}
	\end{subfigure}
	\begin{subfigure}[b]{0.40\textwidth}
		\centering
		\includegraphics[width=\textwidth]{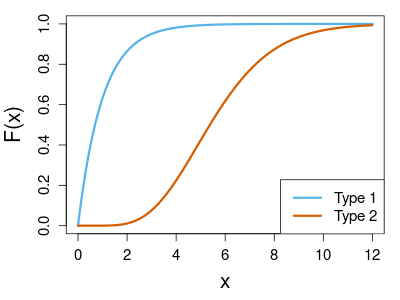}
		%\label{fig:three sin x}
	\end{subfigure}
	\caption{Two model  distributions of Type 1 and Type 2, which are Gamma(1,1) and Gamma(7.5, 0.75) respectively, are visualized by their probability density functions (left) and cumulative distribution functions (right).}
	\label{fig:univariate_pdf_cdf}
\end{figure}

This example aims at comparing how the Wasserstein barycenter and median estimates behave differently in the presence of outliers with a set of univariate distributions.  We consider the Gamma distribution as the model probability measure to which our comparison is applied. In the rest of this experiment, we will denote the Gamma distribution as Gamma($k,\theta$) where $k$ and $\theta$ are shape and scale parameters. We take two distributions, Gamma(1,1) and Gamma(7.5, 0.75), as sources of \textit{signal} (type 1) and \textit{contamination} (type 2), respectively. As shown in Figure \ref{fig:univariate_pdf_cdf}, the two distributions were selected to be largely disparate. To describe, Gamma(1,1) has a monotonically decreasing density function with its mass concentrated near zero while Gamma(7.5, 0.75) has a mound-shaped density. 

Our experiment procedure is as follows. It starts by generating perturbed versions of a distribution by drawing a random sample of varying size $n$ from each distribution, from which an empirical cumulative distribution function (CDF) is computed. This procedure is repeated $(100-k)$ times for Gamma(1,1) and $k$ times for Gamma(7.5, 0.75), leading to a collection of 100 empirical distributions. We set the number of sampling from type 2 measure as $k\in [1,25]$. Since we consider $k$ empirical distributions as contamination out of a total of 100 distributions, we will denote  the degree of contamination as $k\%$. When estimates of the Wasserstein barycenter and median are computed from the collection of 100 empirical CDFs are obtained, the discrepancy between an estimate and the signal - Gamma(1,1) distribution - is measured by the 2-Wasserstein distance. This procedure elaborates how much of an error is induced for a centroid estimator in the presence of outliers.

\begin{figure}[ht]
	\centering
	\begin{subfigure}[b]{0.22\textwidth}
		\centering
		\caption{}
		\includegraphics[width=\textwidth]{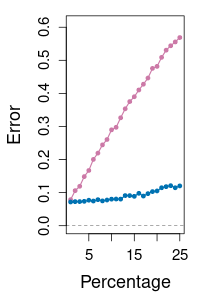}
		%\label{fig:y equals x}
	\end{subfigure}
	\hfill
	\begin{subfigure}[b]{0.22\textwidth}
		\centering
		\caption{}
		\includegraphics[width=\textwidth]{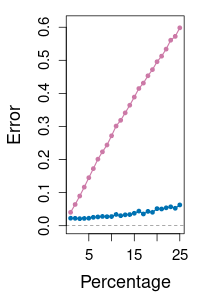}
		%\label{fig:three sin x}
	\end{subfigure}
	\hfill
	\begin{subfigure}[b]{0.22\textwidth}
		\centering
		\caption{}
		\includegraphics[width=\textwidth]{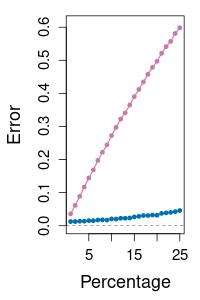}
		%\label{fig:three sin x}
	\end{subfigure}
	\hfill
	\begin{subfigure}[b]{0.22\textwidth}
		\centering
		\caption{}
		\includegraphics[width=\textwidth]{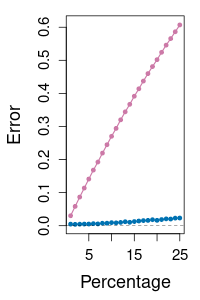}
		%\label{fig:three sin x}
	\end{subfigure}
	\caption{Performance comparison for the univariate distribution example. Average error of 50 runs is measured between the signal distribution and two centroid estimates, the Wasserstein barycenter (in {\color{wass-mean}\bf purple}) and the Wasserstein median (in {\color{wass-meds}\bf blue}), across varying degrees of contamination using the 2-Wasserstein distance. The size of each random sample to reconstruct an empirical distribution is (a) 10, (b) 50, (c) 100, and (d) 500.}
	\label{fig:univariate_performance}
\end{figure}

We performed the experiment with varying sizes of random sample for $n=10, 50, 100$, and $500$ with results summarized  in Figure \ref{fig:univariate_performance}. A universal pattern was observed across different sample sizes the  discrepancy between the barycenter and the signal distribution is magnified as the degree of contamination increases. In contrast, the added contamination does not seem to much affect the Wasserstein median, which shows little deviation from the signal measure.

\begin{figure}[ht]
	\centering
	\begin{subfigure}[b]{0.22\textwidth}
		\centering
		\caption{}
		\includegraphics[width=\textwidth]{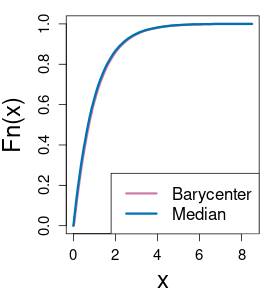}
		%\label{fig:y equals x}
	\end{subfigure}
	\hfill
	\begin{subfigure}[b]{0.22\textwidth}
		\centering
		\caption{}
		\includegraphics[width=\textwidth]{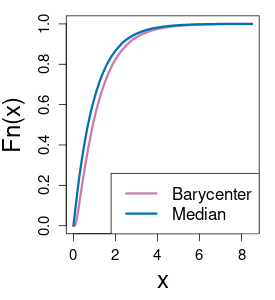}
		%\caption{}
		%\label{fig:three sin x}
	\end{subfigure}
	\hfill
	\begin{subfigure}[b]{0.22\textwidth}
		\centering
		\caption{}
		\includegraphics[width=\textwidth]{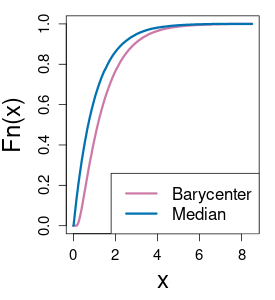}
		%\label{fig:three sin x}
	\end{subfigure}
	\begin{subfigure}[b]{0.22\textwidth}
		\centering
		\caption{}
		\includegraphics[width=\textwidth]{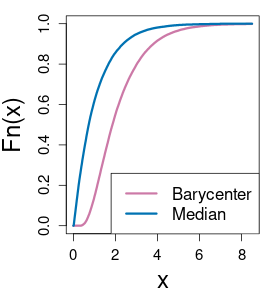}
		%\label{fig:three sin x}
	\end{subfigure}
	\caption{Visualization of the Wasserstein barycenter (in {\color{wass-mean}\bf purple}) and median (in {\color{wass-meds}\bf blue}) from a collection of 100 empirical distributions. Each empirical distribution is constructed from a corresponding random sample of size 500. The degrees of contamination are (a) $1\%$, (b) $5\%$, (c) $10\%$, and (d) $25\%$.}
	\label{fig:univariate_visualization}
\end{figure}

Comparative advantage of the Wasserstein median as a robust centroid is also verified in Figure \ref{fig:univariate_visualization}, where the estimated barycenter and median empirical distributions are visualized across different levels of contamination. When $k$ is small, the two estimates do not much deviate from the cumulative distribution function of the signal as shown in Figure \ref{fig:univariate_pdf_cdf}. As $k$ gets larger, however, the attained barycenter shows a large magnitude  of deviation while the estimated median remains sufficiently close to the cumulative distribution function of the signal. This re-assures the robustness property of the Wasserstein median against the barycenter in the presence of contamination. 

This example and its implication can be understood from a different point of view as well. As noted before, the 2-Wasserstein distance on $\calP_2(\bbR^1)$ is defined as the standard $L_2$ norm between quantile functions of cumulative distribution function. That is, it endows the Hilbertian structure onto the space of probability measures via bijective transformation of probability measures into quantile functions. It was shown that the median in the function spaces is also a robust alternative to the mean \citep{minsker_2015_GeometricMedianRobust}. Hence, the Wasserstein median on $\calP_2(\bbR^1)$ corresponds to a probability measure, the inverse of which is again a median of a collection of quantile functions.

\subsection{Real Example 1 : Newcomb's Speed of Light}

The first real data example involves Newcomb's measurements of the speed of light \citep{newcomb_1891_MeasuresVelocityLight}. The data are composed of 66 scalar-valued measurements collected on different days recorded in millionths of a second. In a version of the dataset where the true speed of light is scaled to $33$, out of 66 observations, two measurements are negative: $-2$ and $-44$. Putting the possibility of time travel aside, it is reasonable to attribute this phenomenon to measurement errors and view these observations as outliers. 

The main goal of our experiment is to represent uncertainty of the measurements using the Gaussian distribution. Our strategy is as follows. First, the dataset is randomly divided into 6 independent subsets, each consisting of 11 observations, ensuring equal cardinality. For each subset, we model the data with a Gaussian distribution, computing the maximum likelihood estimates for location and scale parameters. This procedure creates a collection of 6 Gaussian distributions that are independently fitted for all subsets, based on which we compare the Wasserstein median and the Wasserstein barycenter. From a functional analysis point of view on the space of probability density functions, the mean and median densities are also computed under the standard $L_2$ space structure.

\begin{figure}[ht!]
	\centering
	\begin{subfigure}[b]{0.25\textwidth}
		\centering
		\caption{}
		\includegraphics[width=\textwidth]{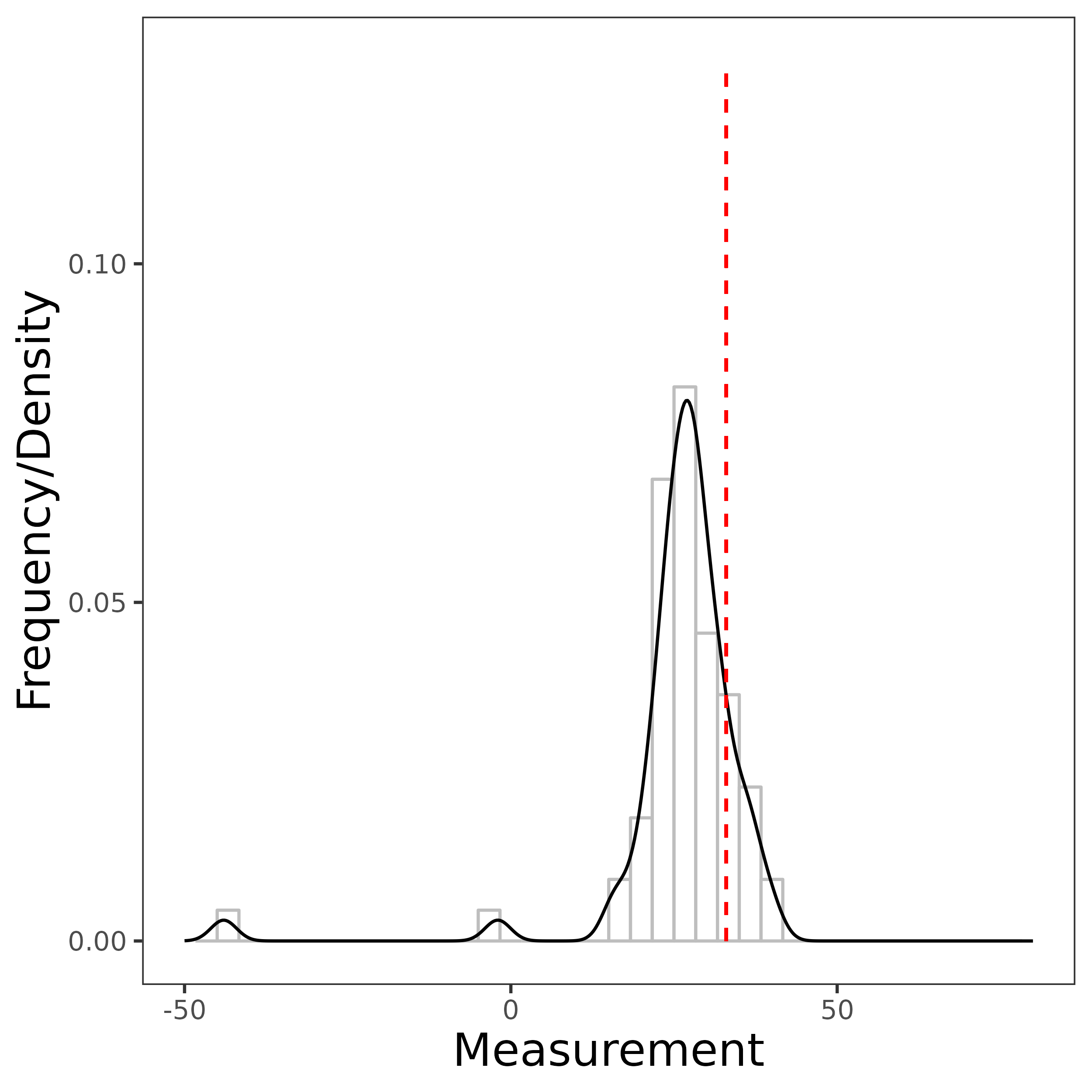}
		%\label{fig:y equals x}
	\end{subfigure}
	\hfill
	\begin{subfigure}[b]{0.25\textwidth}
		\centering
		\caption{}
		\includegraphics[width=\textwidth]{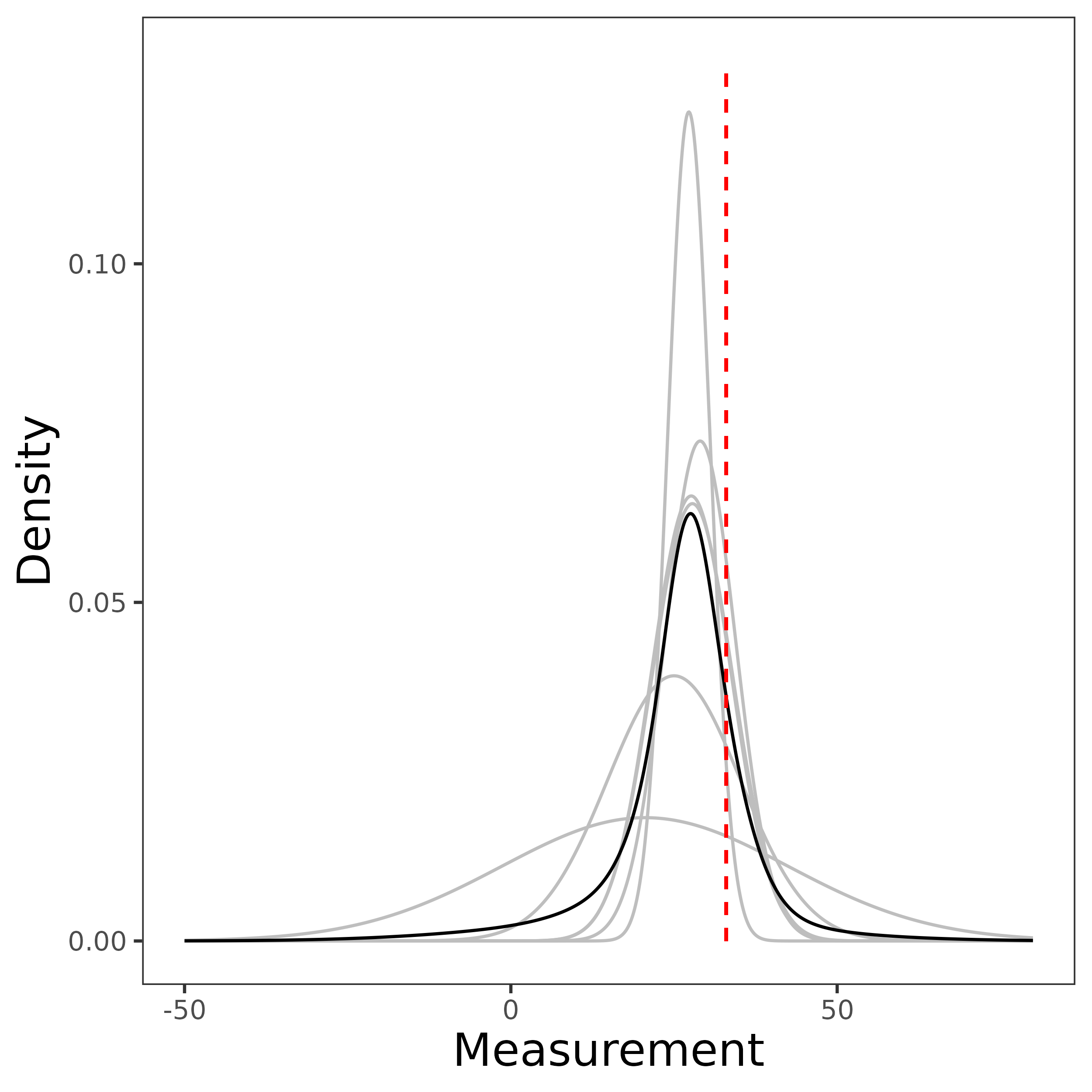}
		%\caption{}
		%\label{fig:three sin x}
	\end{subfigure}
	\hfill
	\begin{subfigure}[b]{0.25\textwidth}
		\centering
		\caption{}
		\includegraphics[width=\textwidth]{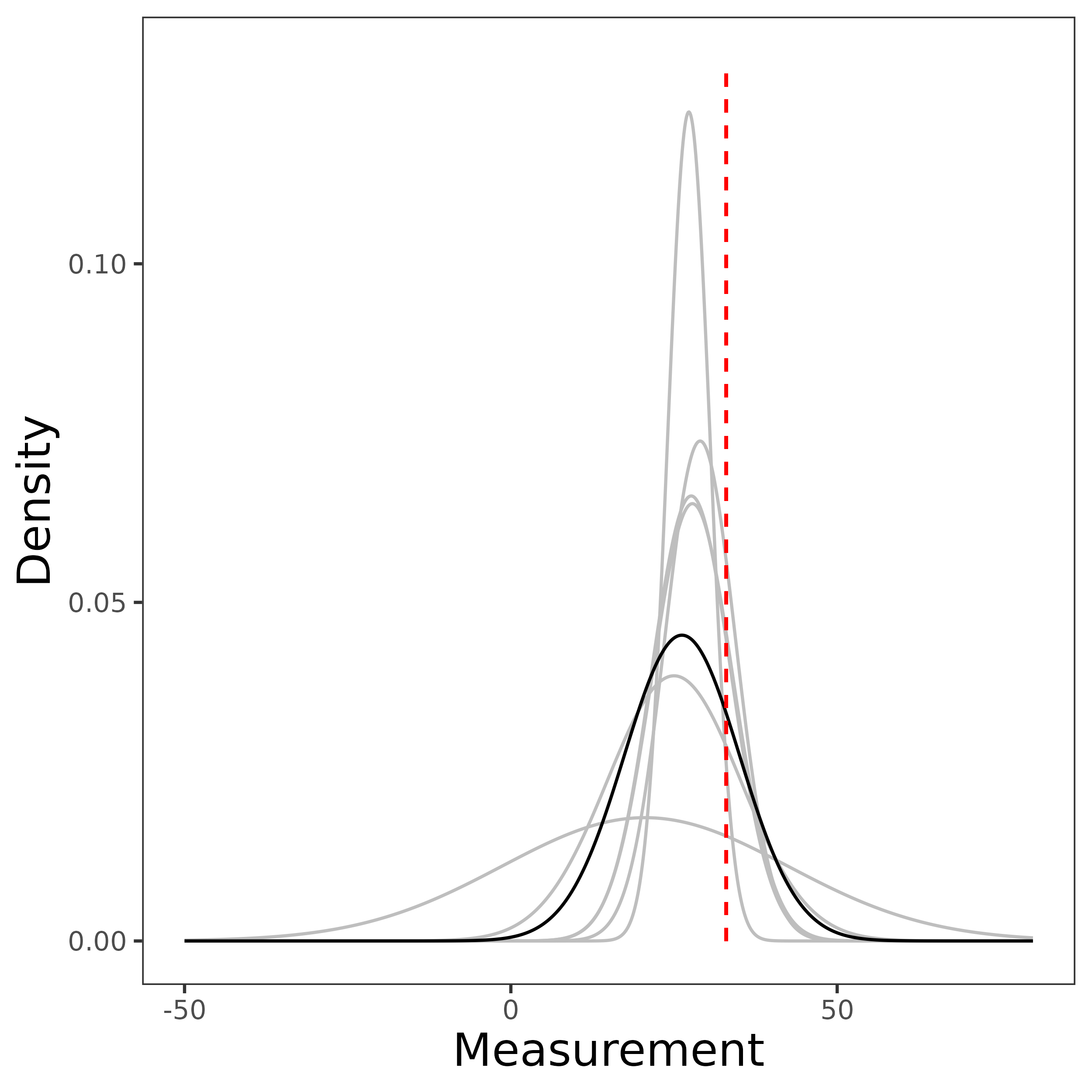}
		%\caption{}
		%\label{fig:three sin x}
	\end{subfigure}
	\hfill
	\begin{subfigure}[b]{0.25\textwidth}
		\centering
		\caption{}
		\includegraphics[width=\textwidth]{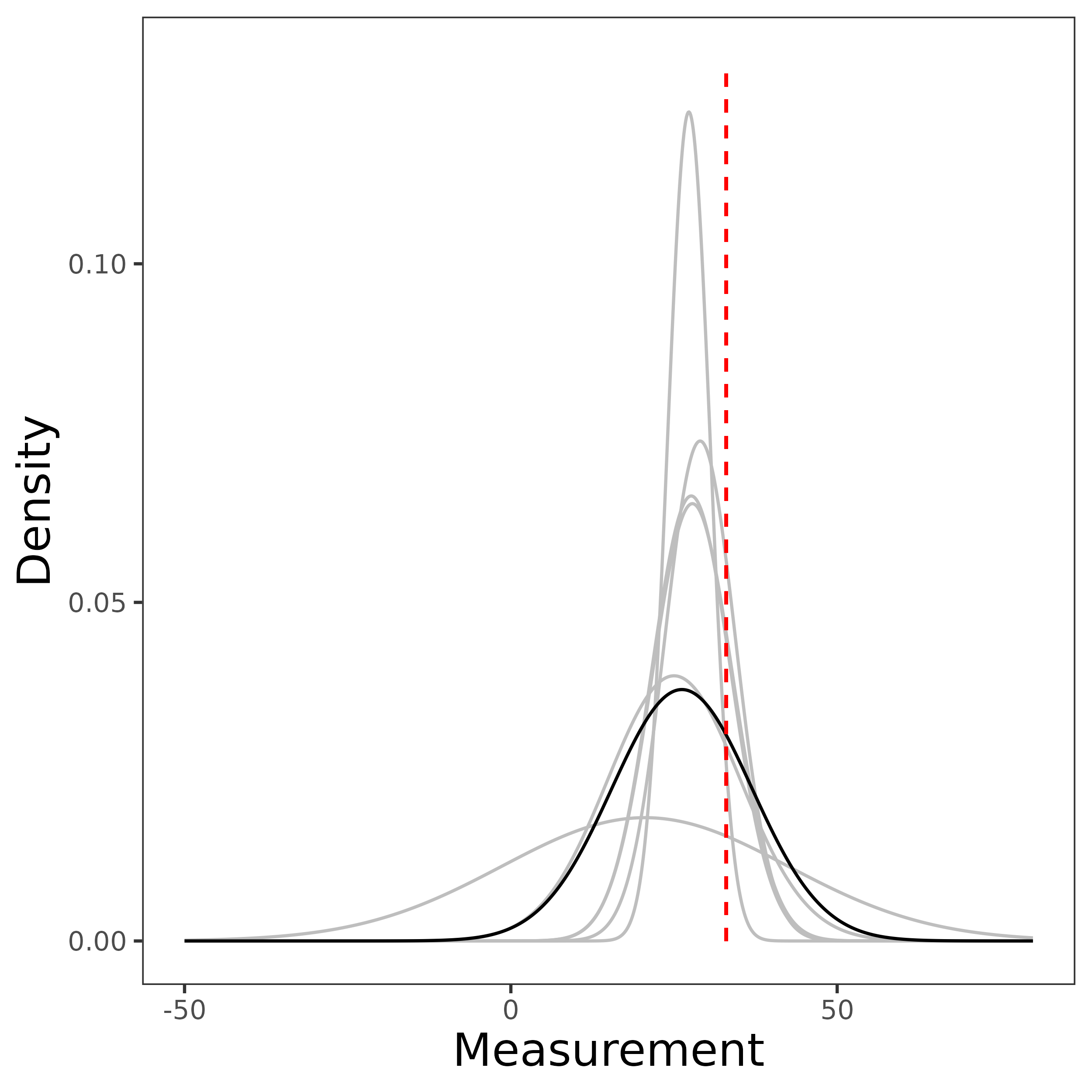}
		%\label{fig:y equals x}
	\end{subfigure}
	\hfill
	\begin{subfigure}[b]{0.25\textwidth}
		\centering
		\caption{}
		\includegraphics[width=\textwidth]{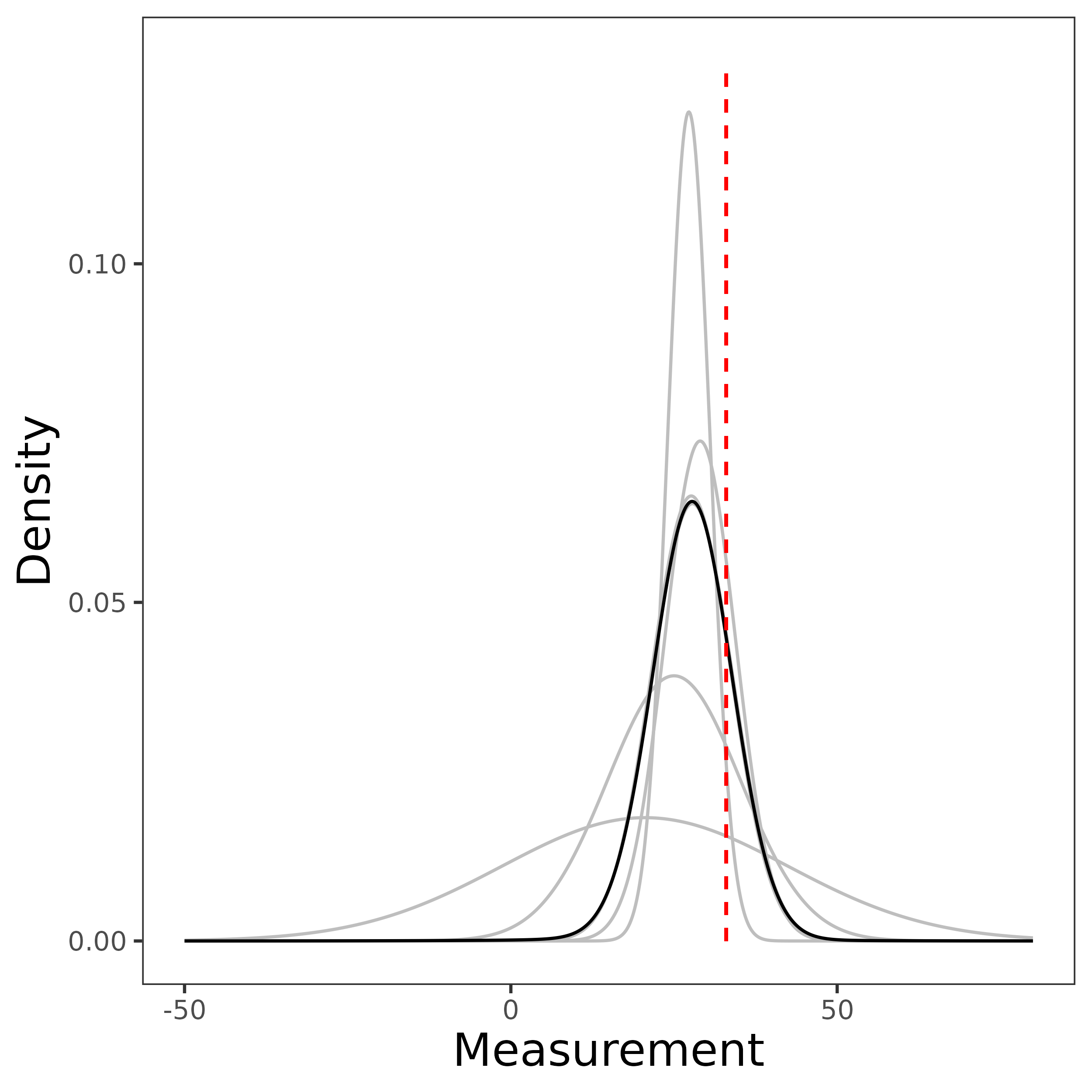}
		%\caption{}
		%\label{fig:three sin x}
	\end{subfigure}
	\hfill
	\begin{subfigure}[b]{0.25\textwidth}
		\centering
		\caption{}
		\includegraphics[width=\textwidth]{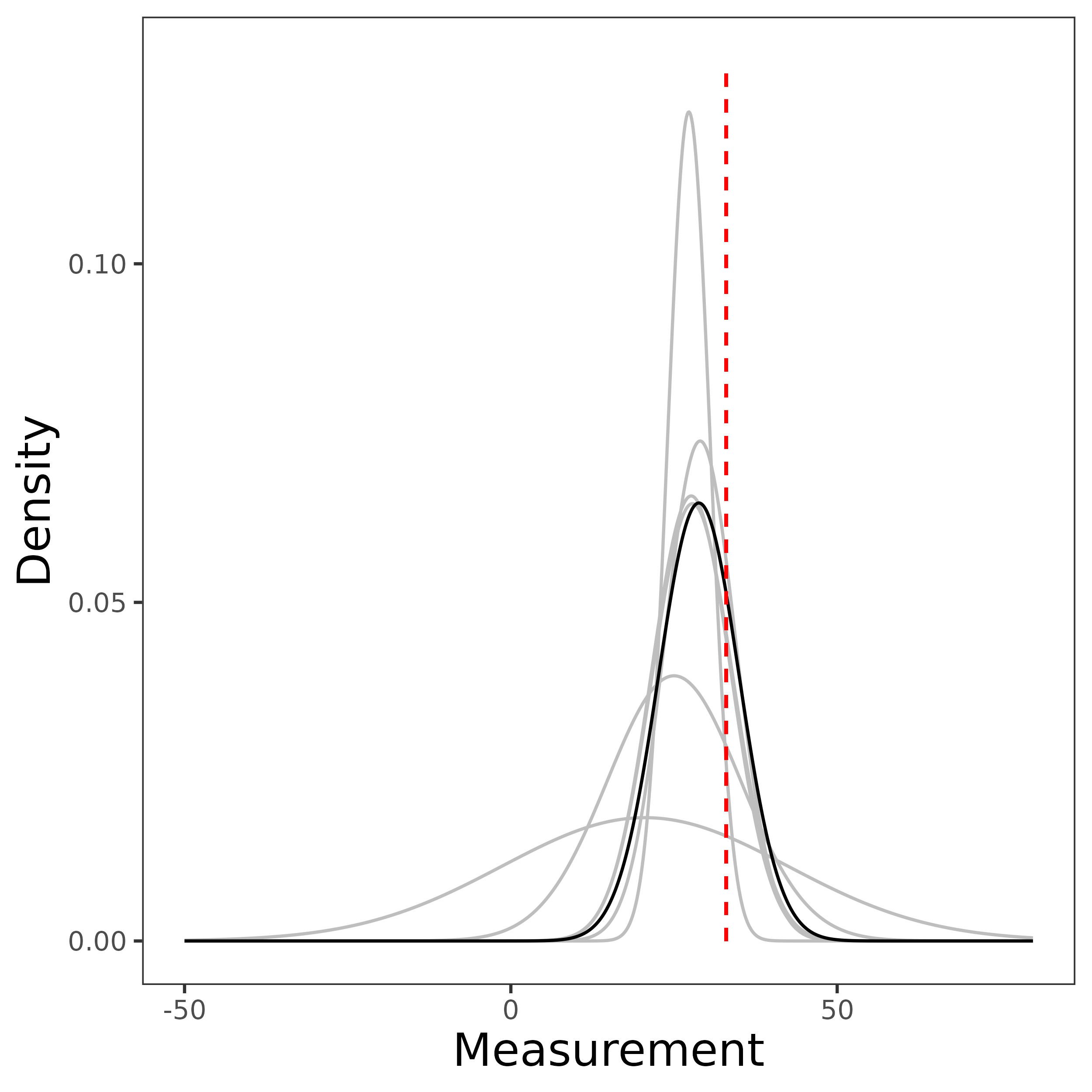}
		%\label{fig:three sin x}
	\end{subfigure}
	\caption{Speed of light measurements example. The left column shows (a) histogram and kernel density estimate of the data and (d) an estimated Gaussian model from the entire data. The center column shows (b) mean and (e) median of subset estimates in $L_2$ sense.The right column presents (c)  barycenter and (f) Wasserstein median of subset estimates. Black  lines represent the estimated densitie while the vertical dashed line in red is the true speed of light. For (b)-(f),  gray lines are density functions of 6 subset estimates.}
	\label{fig:speed_of_light}
\end{figure}

Figure \ref{fig:speed_of_light} presents the aforementioned estimates along with the histogram and kernel density estimate of the data. We note that using a single Gaussian distribution to describe the data and its uncertainty yields the poorest performance among the methods compared.  It is reasonable to expect that the presence of outliers significantly affects the descriptive quality. When we take a look at the divide and conquer approach of multiple subset estimates, it is clear that the Wasserstein median outperforms the barycenter in terms of the final aggregated estimate. The distribution obtained as the Wasserstein median more closely estimates the true speed of light's value. It also exhibits a smaller degree of dispersion than the Wasserstein barycenter, a phenomenon observed in previous examples, indicating a more compact support for the final measure. Unlike the Wasserstein geometry, the mean and median  densities under the standard $L_2$ space structure show little difference with higher concentration of measure that the Wasserstein estimates. Notably, the $L_2$ median estimate is almost identical to the Wasserstein median. In other words, the Wasserstein median can be a compelling alternative to the $L_2$ median in the space of univariate probability measures. We make a remark that this observation is not surprising in the case of $\calP(\bbR^1)$. As noted in our simulated example, the Wasserstein geometry of $\calP(\bbR^1)$ accounts for the Hilbert space structure on the space of quantile functions. Due to the bijection between a measure's density and its distribution functions, the two geometries may induce a dual structure, a property not generally known for higher dimensions, e.g., $\calP(\bbR^d)$ for $d\geq 2$.

\subsection{Real Example 2 : MNIST Digits}

\begin{figure}[ht!]
	\centering
	\begin{subfigure}[]{0.25\textwidth}
		\centering
		\includegraphics[width=\textwidth]{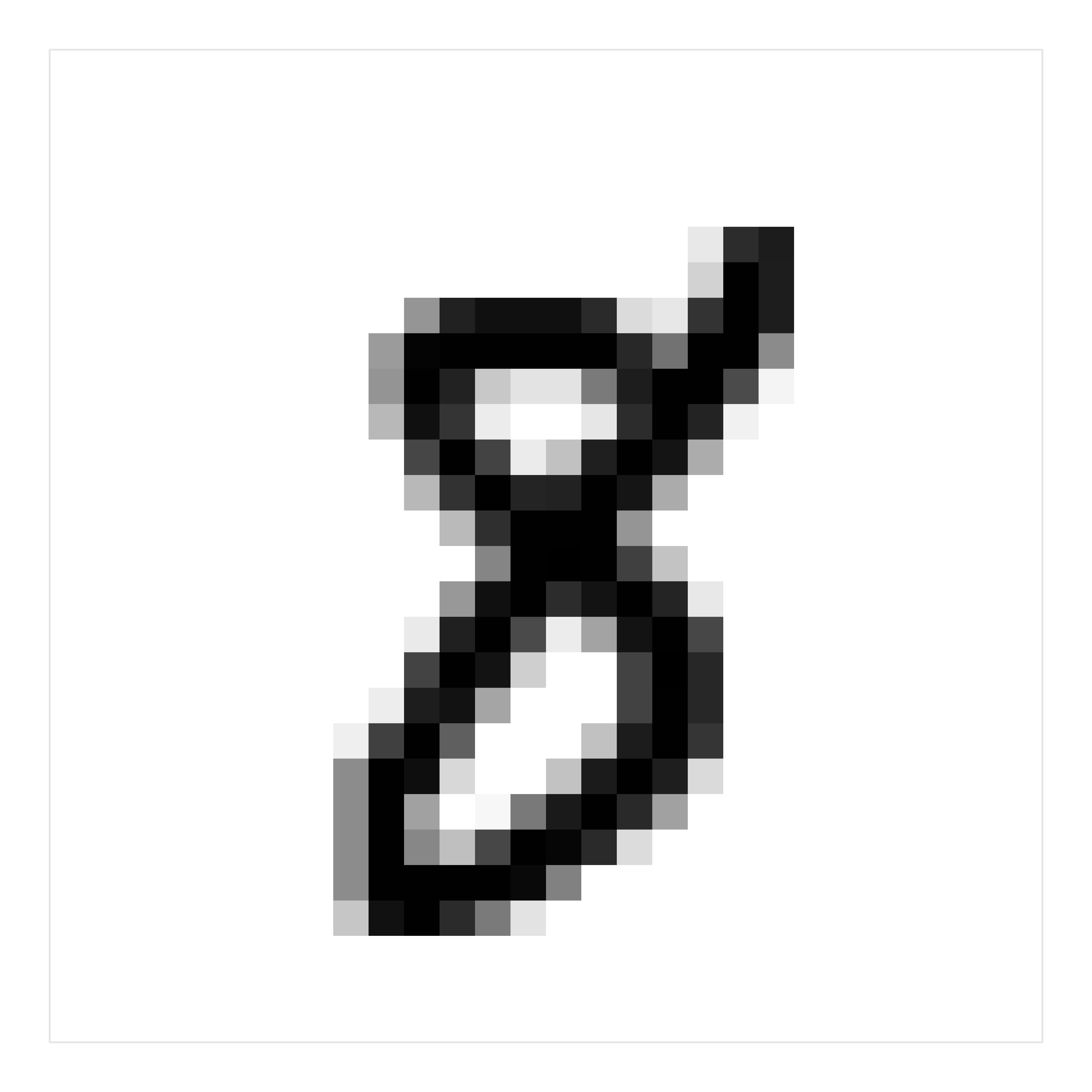}
		%\label{fig:y equals x}
	\end{subfigure}
	\begin{subfigure}[]{0.25\textwidth}
		\centering
		\includegraphics[width=\textwidth]{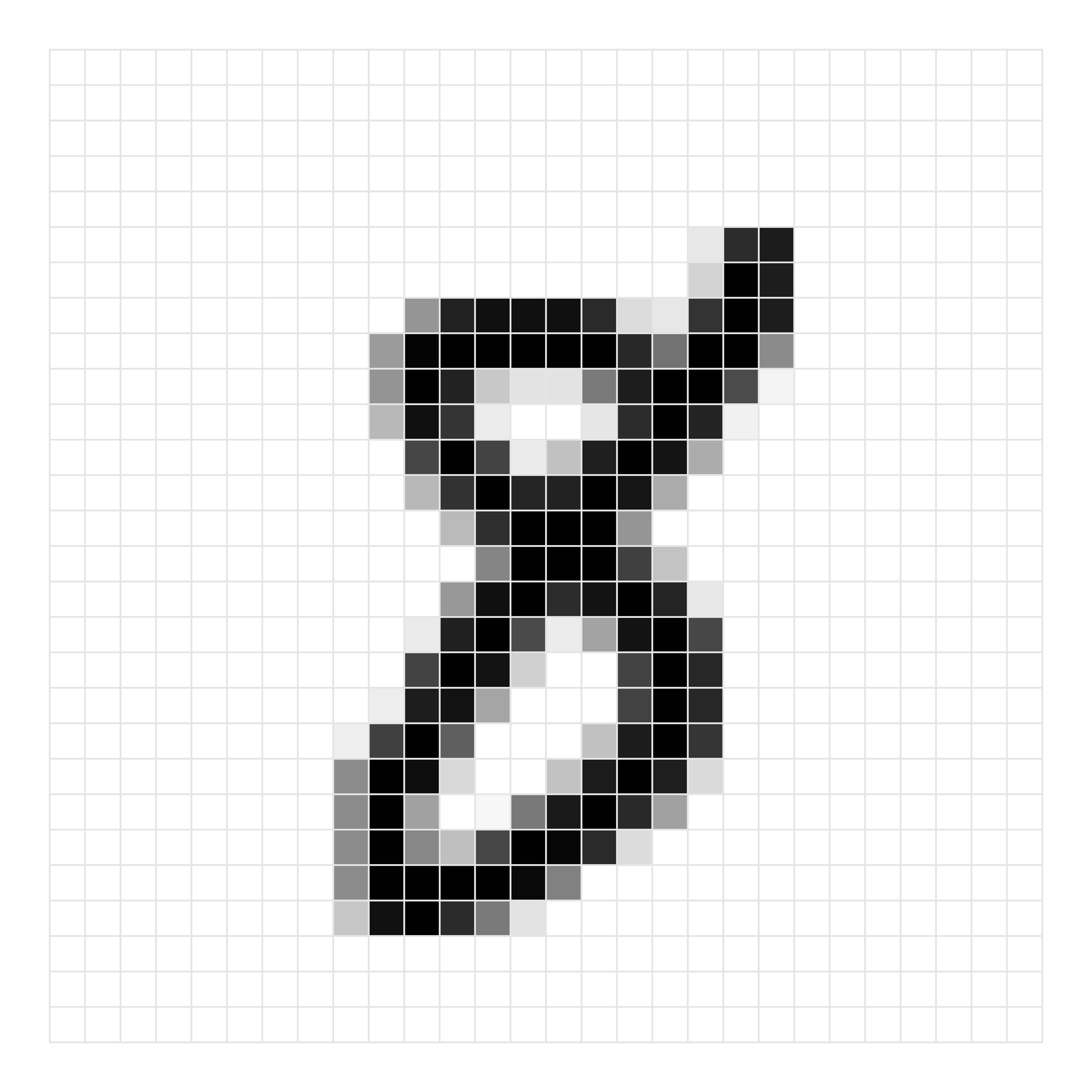}
		%\label{fig:three sin x}
	\end{subfigure}
	\begin{subfigure}[]{0.25\textwidth}
		\centering
		\includegraphics[width=\textwidth]{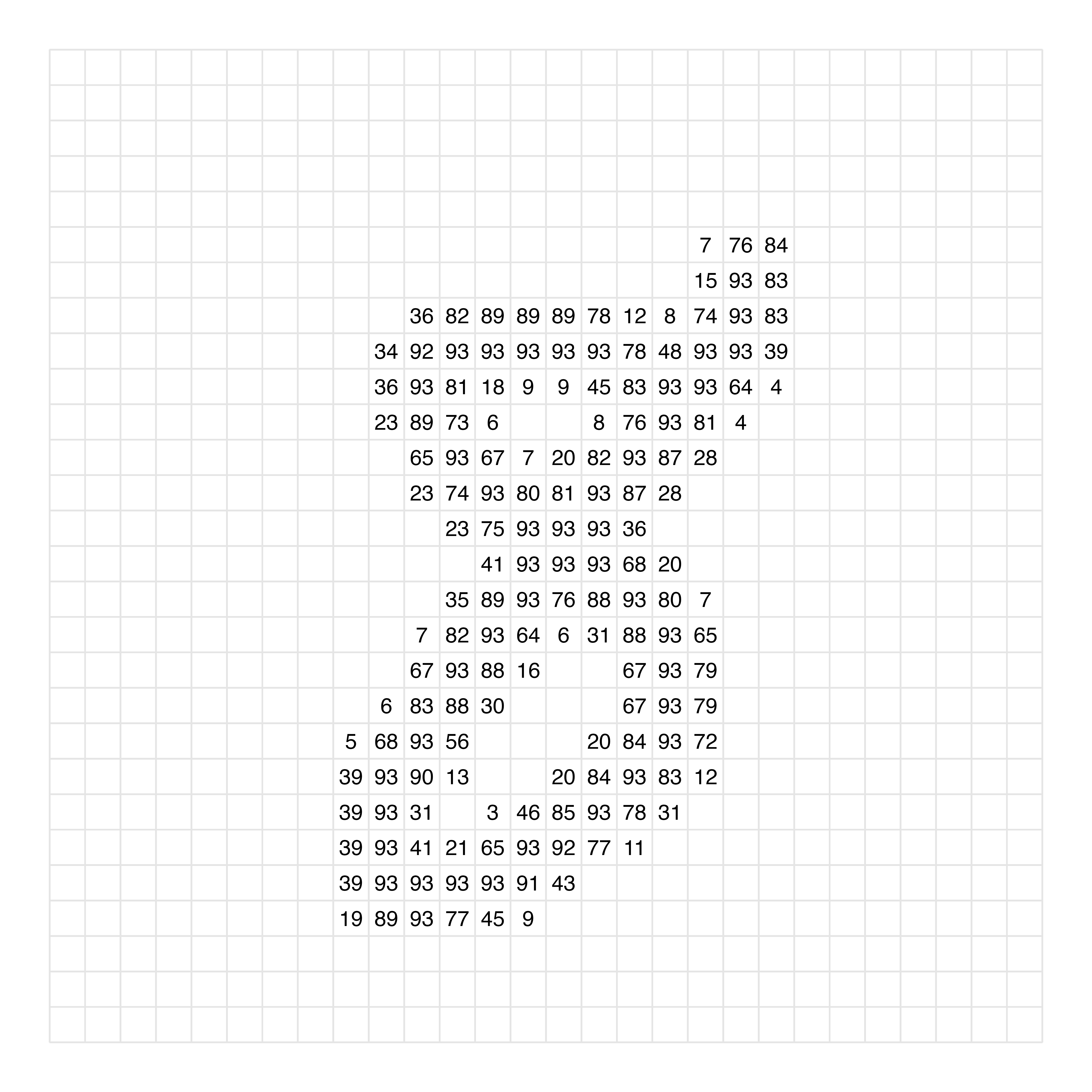}
		%\label{fig:three sin x}
	\end{subfigure}
	\caption{Characterization of a grayscale image as a discrete measure. For an image (left), its regular grid structure encodes location information of each pixel (center). A center point of each pixel represents its location in $\bbR^2$ when convolved with dirac functions. Pixel intensities (right) play a role of weights for point masses to define a discrete measure.}
	\label{fig:image_as_measure}
\end{figure}

The second example is the MNIST handwritten digits, which is a popular dataset in computer vision and image processing \citep{lecun_1998_MNISTDatabaseHandwritten}. The dataset contains grayscale images of handwritten numeric digits from 0 to 9, where each image is represented by a $28\times 28$ matrix and each entry takes a value in $[0,1]$. We applied $L_1$ normalization to each image to ensure that the sum of all pixel values equals 1. Each grayscale image can be viewed as masses distributed along a 2-dimensional regular grid. Hence, a collection of all pixels' locations and intensity values constitutes a discrete probability measure on $\mathbb{R}^2$ as visualized in Figure \ref{fig:image_as_measure}. For each digit, all images were centered and rotated to minimize morphological variations.

\begin{figure}[ht]
	\centering
	\includegraphics[width=.9\linewidth]{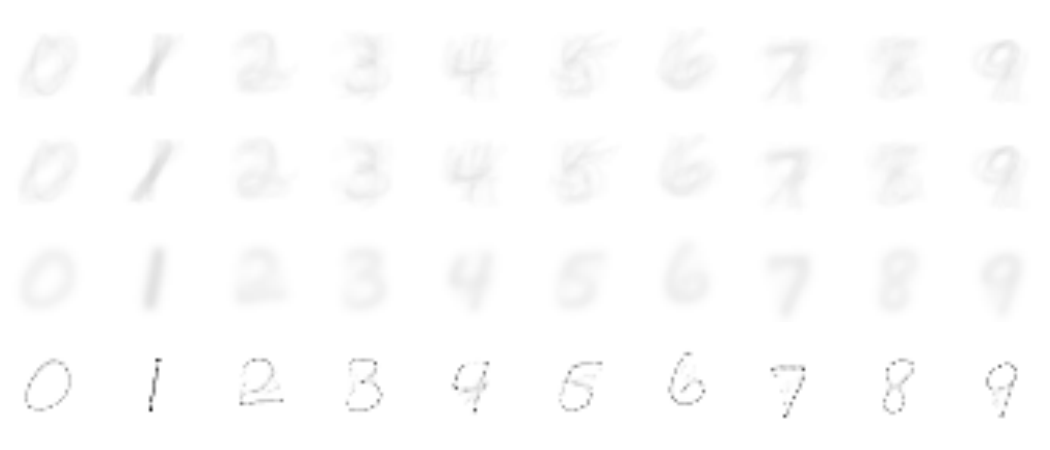}
	\caption{Visualization of various centroids from the MNIST example. Each row represents the Euclidean mean (1st row), the Euclidean median (2nd row), the Wasserstein barycenter (3rd row), and the Wasserstein median (4th row) of 50 images per digit.}
	\label{fig:image_digits}
\end{figure}

Our aim with the MNIST dataset is twofold: visual demonstration of different centroids and classification of new samples based on the estimated centroids. First, we compare different centroids to argue visually that the Wasserstein median possesses appealing characteristics. For each digit, we randomly selected 50 images and computed the centroids, including the Wasserstein barycenter, Wasserstein mean, arithmetic mean, and geometric median. The last two consider images as matrices abiding by the standard Euclidean geometry, and hence we will denote them as Euclidean mean and Euclidean median, respectively. Estimates of the 4 centroids for each digit are presented in Figure \ref{fig:image_digits}. Under Euclidean geometry, two centroids fail to capture the overall shape of each digit, whereas the Wasserstein centroids reveal clear morphological features, producing readily recognizable shapes. This discrepancy arises naturally because the Euclidean mean captures the pixel-wise average of intensity values, and the Euclidean median offers a robust alternative at the pixel level. It is noteworthy that although the Wasserstein centroids surpass their Euclidean counterparts, the Wasserstein median more clearly reveals the morphological characteristics of each digit than the barycenter does. In our grayscale image experiment, the Wasserstein median induces sparse support with more contrastive intensities, bringing higher contrast within an image and better perceptibility for each shape. One explanation to this phenomenon is the robustness of the Wasserstein median against outliers, hence it is more suitable to discover skeletal structure of each digit that functions as basis of our perception.

Next, we performed classification of new samples using a nearest centroid classifier \citep{manning_2008_IntroductionInformationRetrieval}, which assigns each observation to the label of a centroid that attains the minimum distance. After excluding images used for training centroids, we randomly sampled 100 images per digit in that our testing dataset consists a total of 1000 images without class imbalance. Since the Euclidean distance does not respect intrinsic geometry of digits as revealed in our previous demonstration, we used the 2-Wasserstein distance upon which an assignment mechanism is based. Performance of our classification experiment was measured by popular metrics such as accuracy, precision, recall, and F1 score.

\begin{table}[]
	\centering
	\begin{tabular}{|c|cccc|}
		\hline
		\multirow{2}{*}{Centroids} & \multicolumn{4}{c|}{Performance Metrics}                                                              \\ \cline{2-5} 
		& \multicolumn{1}{c|}{Accuracy} & \multicolumn{1}{c|}{Precision} & \multicolumn{1}{c|}{Recall} & F1-Score \\ \hline		
		Euclidean mean     & \multicolumn{1}{c|}{0.583} & \multicolumn{1}{c|}{0.612}   & \multicolumn{1}{c|}{0.583}       &      0.597   \\ \hline
		Euclidean median     & \multicolumn{1}{c|}{0.561} & \multicolumn{1}{c|}{0.575}   & \multicolumn{1}{c|}{0.561}       &      0.567    \\ \hline
		Wasserstein barycenter     & \multicolumn{1}{c|}{0.692} & \multicolumn{1}{c|}{0.698}   & \multicolumn{1}{c|}{ 0.692}       &      0.695   \\ \hline
		Wasserstein median     & \multicolumn{1}{c|}{0.735} & \multicolumn{1}{c|}{0.746}   & \multicolumn{1}{c|}{0.735 }       &      0.740  \\ \hline
	\end{tabular}
	\caption{MNIST classification results using a nearest centroid classifier for a balanced test set of 1000 images across 10 digit classes. All centroids were estimated from 50 images per digit.}
	\label{table:classification}
\end{table}

According to Table \ref{table:classification} that summarizes performance metrics, we observe the similar pattern as before that the Wasserstein centroids are superior to the Euclidean counterparts. Again, it is highlighted that the Wasserstein median significantly outperforms the barycenter in a nearest centroid classifier across all metrics. Specifically, the Wasserstein median has higher precision and recall, which is ideal in a sense that it accurately identified the majority of actual positives with minimal false positives. This observation again highlights an advantage of the Wasserstein median to capture foundational structure of each digit, providing higher degree of generalizability and predictability for perturbed images.

\subsection{Real Example 3 : Cluster Analysis of Population Pyramids}

A population pyramid, also known as age-sex pyramid, is an informative representation of the distribution of a population by age groups and sex \citep{preston_2009_DemographyMeasuringModeling}. The literature often considers trichotomous categorization of population pymarids according to the shapes of their distributions: expansive, constrictive, and stationary.

An expansive pyramid represents an abundance of the younger population and narrows with increasing age, often indicative of a high birth rate and short life expectancy. A constrictive pyramid features a narrow base for younger ages and a larger mass at higher ages. It is often understood as an indication of high-quality healthcare, few adverse environmental factors, and a low birth rate. The last category, stationary, is where almost all age groups occupy nearly constant percentages.

\begin{figure}[ht]
	\centering
	\includegraphics[width=.9\linewidth]{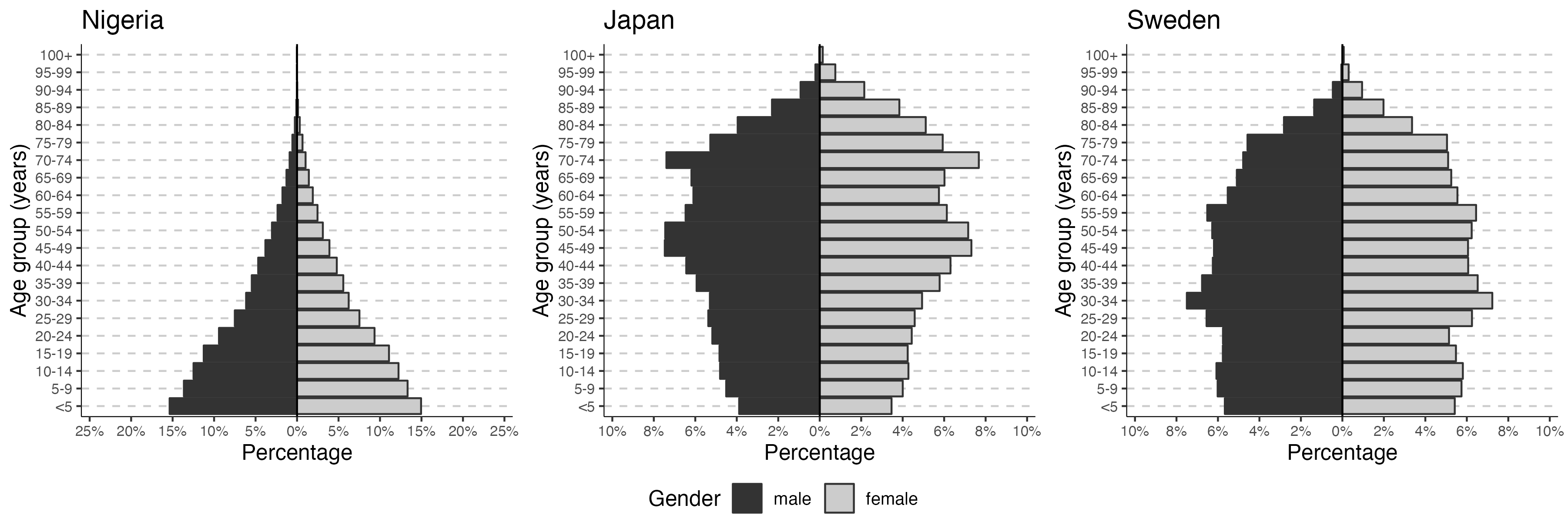}
	\caption{Exampler population pyramids according to the trichotomous categorizations of expansive (left; Nigeria), constrictive (middle; Japan), and stationary (right; Sweden) classes. Population pyramids are visualized by percentage of each age group rather than the population counts to focus on compositional feature.}
	\label{fig:real3-representative}
\end{figure}

Our goal is to validate the aforementioned hypothesis on categorization of population pyramids with cluster analysis. For the real data analysis, we downloaded the International Database (IDB) maintained by the U.S. Census from its website\footnote{\url{https://www.census.gov/programs-surveys/international-programs/about/idb.html}}. The IDB dataset compiles demographic measures at the national and subnational levels, including estimation of population sizes since the 1960s. We extracted a temporal snapshot of national-level population counts in the year of 2022, which is the most recent year with complete data availability according to the data description. From the real data, we visualized population pyramids of some of the most typical countries for the three categories - Nigeria (expansive), Japan (constrictive), and Sweden (stationary), as shown in Figure \ref{fig:real3-representative}, where the difference between constrictive and stationary shapes is not as prominent as with the expansive type.

For data-driven discovery of representative population pyramid types, we applied the $k$-partitional clustering algorithm \citep{sammut_2011_EncyclopediaMachineLearning}, which decomposes a dataset into disjoint clusters for some fixed $k$ that accounts for the number of clusters. 

If an observation is assigned to a class based on its proximity to the centroid of its corresponding cluster, the $k$-partitional clustering can be considered a meta-algorithm that includes $k$-means \citep{macqueen_1967_MethodsClassificationAnalysis}, $k$-medoids \citep{kaufman_1990_PartitioningMedoidsProgram}, among others, as its special cases. We compared the effectiveness of different centroids, including the mean and median, under both Euclidean and Wasserstein geometries, within the partitional cluster analysis framework. The number of clusters $k$ was set to vary from 2 to 10 in order to examine varying community structures at different scales. Once clustering is complete, we measured validity of the attained partition structure using the silhouette score \citep{rousseeuw_1987_SilhouettesGraphicalAid}, which measures the relative expression of cohesive patterns to the degree of separation induced by a partition. We note that a silhouette score is computed for each observation, hence we used the average of all silhouette values to determine overall quality of a partition.

\begin{figure}[ht]
	\centering
	\includegraphics[width=.95\linewidth]{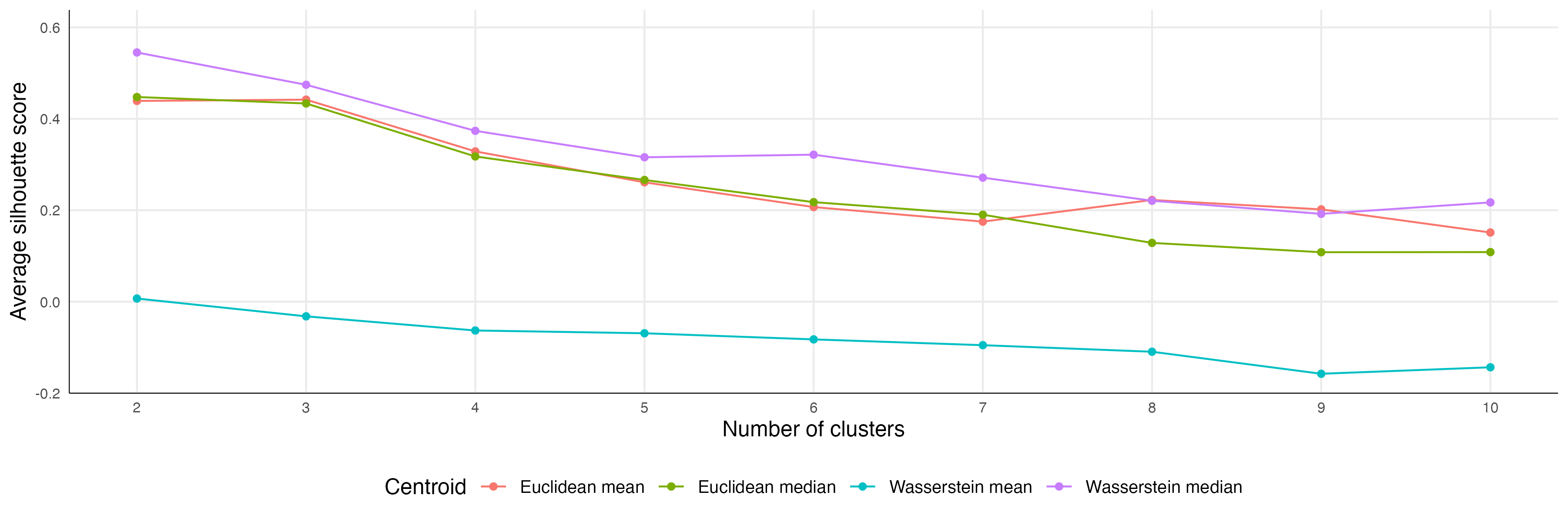}
	\caption{Average silhouette scores for cluster labels obtained by the $k$-partitional clustering algorithm with the varying numbers of clusters $k=2$ to $10$ and  4 different types of centroids considered.}
	\label{fig:real3-silhouette}
\end{figure}

We summarized performance of the partitional clustering algorithm in Figure \ref{fig:real3-silhouette}, examining various $k$ values and types of centroids.

For all centroid types, $k=2$ exhibited the highest average silhouette score, signifying superior clustering quality. This implies that two distinct categories of population pyramids were discovered in a data-driven manner. This alignment underscores the question regarding the distinguishability between constrictive and stationary pyramid types. Although heterogeneous social dynamics may define the qualitative boundaries between them, differentiating the two quantitatively is less straightforward. This reasoning is also consistent with visual inspection on the results of cluster analysis.

\begin{figure}[ht]
	\centering
	\includegraphics[width=.95\linewidth]{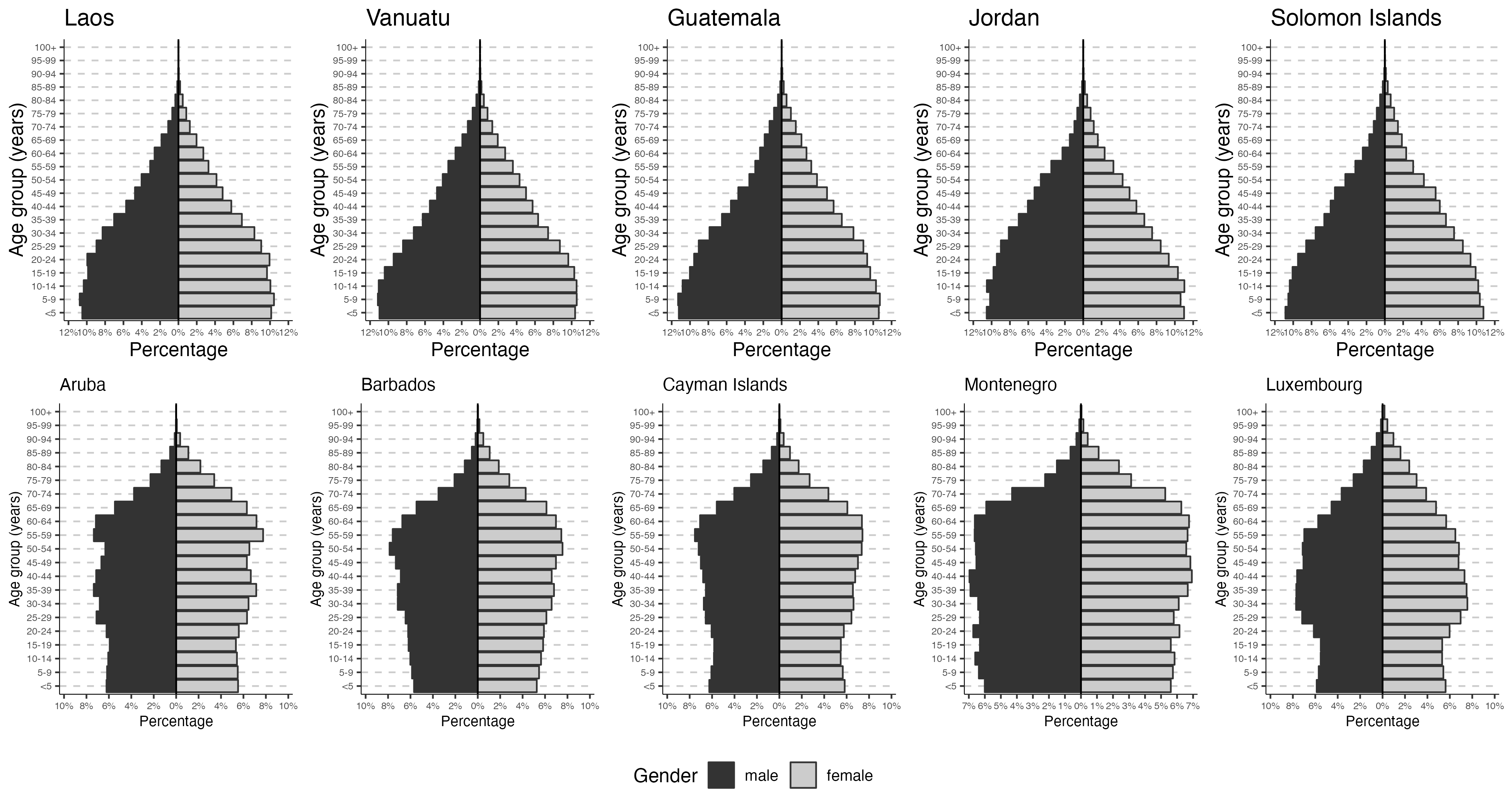}
	\caption{Population pyramids of 5 countries that are nearest to the Wasserstein median centroid from the $k$-partitional clustering algorithm with $k=2$. The two rows correspond to countries in cluster 1 (top) and cluster 2 (bottom), respectively representing the expansive type and a mixture of constrictive and stationary types.}
	\label{fig:real3-countries}
\end{figure}

Figure \ref{fig:real3-countries} shows five countries closest to the centroid of the Wasserstein median for the two classes obtained. Countries in class 1, including Laos, Vanuatu, Guatemala, Jordan, and the Solomon Islands, display the stereotypical shape associated with the expansive pyramid type. Conversely, the second group comprises countries with pyramids that are a mixture of constrictive and stationary types, or somewhere in between.

\section{Conclusion}\label{sec:conclusion}

We proposed the Wasserstein median as a generalization of the geometric median from Euclidean space to the space of probability measures, utilizing the Wasserstein distance of order 2. We presented theoretical results on the existence, consistency, and robustness of the Wasserstein median. A generic algorithm was also proposed that provides a theoretical guarantee of convergence and brings about the uniqueness of the attained median. This algorithm circumvents the repeated estimation of optimal transport maps, a task that is computationally intensive in realistic scenarios. Moreover, our computational pipeline is readily adaptable to any context where a routine for computing the Wasserstein barycenter exists. Through a series of experiments across various domains where optimal transport (OT) tools have been applied, we provided strong empirical evidence supporting the robustness of the Wasserstein median compared to the Wasserstein barycenter, especially in the presence of outliers. We believe our framework not only addresses a fundamental gap in learning with probability measures within the OT framework but also establishes a foundation for its application across a wide range of problems.

In concluding this paper, we outline a brief agenda for future research. A primary focus should be on determining the conditions under which the population version of the Wasserstein median is unique. Our current analysis, limited to finite sample cases, draws from the uniqueness of the Wasserstein barycenter. A critical avenue for exploration involves alterantives to the convexity at the minima, given the Wasserstein space's lack of local compactness. Further investigation into other robustness measures and their manifestation in the Wasserstein median is crucial, as theoretical advancements in this area could significantly bolster its utility. Additionally, exploring the generalization of the base space presents an exciting direction. Our discussion has been confined to Euclidean spaces with absolutely continuous input measures. However, ongoing efforts in the OT literature to extend foundational theories to more general settings, such as locally compact geodesic spaces or nonnegative or nonpositive curved spaces, are noteworthy. Given the frequent necessity to handle data with complex structures in modern statistics, it is imperative to address scenarios where the assumptions of well-established theories are not met. Investigating the asymptotic behavior of the estimator, including convergence rates, is also of interest. Finally, we aim to develop an enhanced computational framework for the Wasserstein median. Our current method, which iteratively utilizes algorithms designed for estimating Wasserstein barycenters, cannot avoid increased computational complexity. A potential solution may lie in a stochastic algorithm tailored to directly address the problem. Developing efficient computational strategies will undoubtedly contribute to the wider adoption of the proposed method.

\appendix
\section*{Appendices}
\subsection*{A. Proof of Theoretical Results}
\subsubsection*{Proposition 1}
\begin{proof}
	If the functional $F(\nu) = \bbE \calW_2 (\nu, \Lambda)$ is identically infinite, the statement is valid for any choice of measure in $\calP_2(\Rd)$. If not, take a minimizing sequence $\lbrace \nu_j\rbrace \subset  \calP_2(\Rd)$ and denote an upper bound of the functional with respect to the sequence as $C = \underset{j}{\sup}~ F(\nu_j) < \infty$. Assume $\nu_0$ be a fixed reference measure. By the triangle inequality, for a Dirac measure $\delta_0$ in $\calP_2(\Rd)$, we have
	\begin{eqnarray*}
		\bbE \calW_2 (\delta_0, \Lambda) &\leq& \bbE \calW_2 (\delta_0, \nu_0) + \bbE \calW_2 (\nu_0, \Lambda)\\
		&=& \calW_2 (\delta_0, \nu_0) + \bbE \calW_2 (\nu_0, \Lambda) \\
		&\leq& \calW_2 (\delta_0, \nu_0) + C.
	\end{eqnarray*}
	Therefore, the following holds for all $j$,
	\begin{eqnarray*}
		\bbE \calW_2 (\nu_j, \delta_0) &\leq& \bbE \calW_2 (\nu_j, \Lambda) + \bbE \calW_2 (\Lambda, \delta_0)\\
		&\leq& C + \calW_2 (\delta_0, \nu_0) + C\\ &=& 2C + \calW_2 (\delta_0, \nu_0) \\ &=& 2C + \left\lbrack \int_{\Rd} \| {\bf  x}\|^2 d\nu_0({\bf x}) \right\rbrack^{1/2} < \infty,
	\end{eqnarray*}
	and denote $\sup_j \bbE \calW_2 (\nu_j, \delta_0) = M < \infty$. For some $R>0$, the Chebyshev's inequality states that
	\begin{equation*}
	\nu_j \left(\left\lbrace {\bf x}\in\Rd ~|~\|{\bf x}\|\geq R \right\rbrace\right) \leq 
	\frac{1}{R^2} \int_{\lbrace {\bf x}:\|{\bf x}\|\geq R\rbrace} \|{\bf x}\|^2 d\nu_n ({\bf x}) \leq \left(\frac{M}{R}\right)^2.
	\end{equation*}
	Take a sufficiently large $R$ such that $R > M^2 / \sqrt{\epsilon}$ for a small $\epsilon > 0$. Then, for a closed and bounded ball $B_{R} = \lbrace {\bf x}\in\Rd ~|~\|{\bf x}\|\leq R \rbrace$, we have
	\begin{equation*}
	\nu_j (B_R) = 1 - \nu_j(B_R^\mathsf{c}) > 1 - \epsilon \quad \text{for all }j,
	\end{equation*}
	which completes to show that sequence of measures $\lbrace \nu_j \rbrace$ is tight by the Heine-Borel theorem.
	
	Since the sequence is tight, there exists a subsequence $\lbrace \nu_{j_k}\rbrace$ that converges weakly to some $\nu^* \in \calP_2 (\Rd)$, which is characterized as a minimizer of the function since
	\begin{equation*}
	F(\nu^*) = \bbE \calW_2 (\nu^*, \Lambda) \leq \bbE ~ \underset{j_k \rightarrow \infty}{\liminf} ~\calW_2 (\nu_{j_k},\Lambda) \leq \underset{j_k \rightarrow \infty}{\liminf} ~\bbE \calW_2 (\nu_{j_k},\Lambda) = \inf F,
	\end{equation*}
	where the first and second inequalities are by Fatou's lemma and lower semicontinuity of the Wasserstein distance. This completes the proof. 
\end{proof}

\subsubsection*{Proposition 2}
\begin{proof}
	We rephrase the three-step strategy as in \cite{legouic_2017_ExistenceConsistencyWasserstein}. The first step is to show that a sequence of Wasserstein medians $\lbrace \nu_j\rbrace$ is tight, which already appeared in the proof of Proposition 1. Therefore, we can take a subsequence $\lbrace \nu_{j_k}\rbrace \subset \lbrace \nu_j \rbrace$ that converges to some $\nu$. 
	
	Next, we show that the accumulation point $\nu$ is also a Wasserstein median. Take some $\xi \in \calP_2(\bbR^d)$ and a random measure $\mu \sim \Lambda$. Note that we employ a slightly different notation in what follows and this is for the purpose of distinguishing the first and the second degree Wasserstein spaces. Our goal is to show that $\bbE \calW_2 (\xi, \mu) \geq \bbE \calW_2 (\nu, \mu)$ for any choice of $\xi$ by similar arguments as before. 
	\begin{align}
	\bbE	\calW_2 (\xi, \mu) &= \calW_2 (\delta_\xi, \Lambda) \nonumber\\
	&=   \lim_{j_k \rightarrow \infty} \calW_2 (\delta_\xi, \Lambda_{j_k}) \nonumber\\
	&= \lim_{j_k \rightarrow \infty} \bbE \calW_2 (\xi, \mu_{j_k})\text{ for }\mu_{j_k} \sim \Lambda_{j_k}\nonumber\\
	&\geq \lim_{j_k \rightarrow \infty} \bbE \calW_2 (\nu_{j_k}, \mu_{j_k}) \label{proof-consistency-firstinequality}\\
	&\geq \bbE \liminf_{j_k \rightarrow \infty} \calW_2 (\nu_{j_k}, \mu_{j_k})  \text{ by Fatou's lemma}\nonumber\\
	&\geq \bbE \calW_2 (\nu, \mu)\label{proof-consistency-lastinequality}.
	\end{align}
	The first inequality \eqref{proof-consistency-firstinequality} holds because every element in $\lbrace \nu_{j_k}\rbrace$ is a Wasserstein median and the last inequality \eqref{proof-consistency-lastinequality} is induced by the fact that Wasserstein distance is lower semicontinuous. Also, convergence of $\Lambda_{j_k} \rightarrow \Lambda$ enables to construct $\mu_{j_k} \rightarrow \mu$ almost surely by the Skorokhod's representation theorem  \citep{billingsley_1999_ConvergenceProbabilityMeasures}. Hence, we assert that any limit point $\nu$ of $\lbrace \nu_j \rbrace$ is indeed a Wasserstein median. 
	
	The last component is to show how $\nu$ is related to a limiting measure $\Lambda$. One implication of \eqref{proof-consistency-lastinequality} is that we have $\calW_2 (\delta_{\nu_{j_k}}, \Lambda_{j_k}) \rightarrow \calW_2 (\delta_\nu, \Lambda)$ if $\xi = \nu$. By the triangular inequality, we have
	\begin{equation*}
	\calW_2 (\delta_{\nu_{j_k}}, \Lambda) - \calW_2 (\delta_{\nu}, \Lambda) \leq \calW_2 (\delta_{\nu_{j_k}}, \Lambda_{j_k}) + \calW_2 (\Lambda_{j_k}, \Lambda) - \calW_2 (\delta_{\nu}, \Lambda). 
	\end{equation*}
	A repeated use of the previous tricks leads to 
	\begin{equation*}
	\liminf_{j_k \rightarrow \infty} \calW_2 (\nu_j, \mu) = \calW_2 (\nu, \mu),
	\end{equation*}
	which implies $\calW_2 (\nu_{j_k}, \xi) \rightarrow \calW_2 (\nu, \xi)$ for any $\xi \in \calP_2 (\bbR^d)$ almost surely with respect to $\Lambda$. By the Theorem 2.2.1 of \cite{panaretos_2020_InvitationStatisticsWasserstein}, this is equivalent to $\calW_2 (\nu_{j_k}, \nu) \rightarrow 0$, which completes the proof. 
\end{proof}

\subsubsection*{Theorem 3}
This is a direct application of Theorem 2 from \cite{fletcher_2004_PrincipalGeodesicAnalysis}, where the breakdown point $\epsilon^*(T,X)$ for a random sample $X$ on an arbitrary Riemannian manifold was demonstrated. The Wasserstein distance allows us to view $\calP_2(\mathbb{R}^d)$ as a Riemannian manifold of non-negative curvature and as a complete metric space. The only component in the proof that utilizes the geometric properties of the underlying space is the triangle inequality, which is guaranteed for the 2-Wasserstein distance; hence, no extra steps are required.

\subsubsection*{Theorem 4}
\begin{proof}
	We first show that the updating map $U:\calP_2 (\bbR^d) \rightarrow \calP_2 (\bbR^d)$ is a continuous function, which is defined by a suitable minimization problem in the IRLS formulation,
	\begin{equation*}
	U(\tilde{\nu}) = \argmin_{\nu} \sum_{n=1}^N \frac{\pi_n}{\calW_2 (\tilde{\nu}, \mu_n)} \calW_2^2 (\nu, \mu_n).
	\end{equation*}
	Denote $\nu_1, \nu_2$ be two arbitrarily close measures, i.e., $\calW_2(\nu_1, \nu_2) <\delta$. By the triangle inequality, we have 
	\begin{equation*}
	\begin{gathered}
	\calW_2 (\nu_1, \mu_n) \leq \calW_2 (\nu_1, \nu_2) + \calW_2 (\nu_2, \mu_n) < \delta + \calW_2(\nu_2, \mu_n), \\ 
	\calW_2 (\nu_2, \mu_n) \leq \calW_2 (\nu_2, \nu_1) + \calW_2 (\nu_1, \mu_n) < \delta + \calW_2(\nu_1, \mu_n),
	\end{gathered}
	\end{equation*}
	so that the inequality $\vert \calW_2(\nu_1, \mu_n) - \calW_2(\nu_2, \mu_n)\vert<\delta$ holds for all $n=1,\ldots,N$. Let $a_n = \calW_2(\nu_1, \mu_n)$ and $b_n = \calW_2 (\nu_2, \mu_n)$. When we consider the following maps
	\begin{equation*}
	U(\nu_1) = \argmin_{\nu} \sum_{n=1}^N \frac{\pi_n}{a_n} \calW_2^2 (\nu, \mu_n),\quad					U(\nu_2) = \argmin_{\nu} \sum_{n=1}^N \frac{\pi_n}{b_n} \calW_2^2 (\nu, \mu_n),
	\end{equation*}
	the last inequality implies that one can control coefficients (or scaled weights) of the two updating maps to be sufficiently close by the following observation
	\begin{equation*}
	\left\vert \frac{\pi_n}{a_n} - \frac{\pi_n}{b_n} \right\vert = \frac{\pi_n}{a_n b_n} \left\vert a_n - b_n \right\vert < \frac{\delta \pi_n}{a_n b_n} .
	\end{equation*}
	This further implies that the smaller the $\delta$ is, the closer two objective functions in the updating maps are. For any absolutely continuous measure $\mu$, the map $\nu \mapsto \calW_2^2 (\nu, \mu)$ is strictly convex and the solutions of two minimization problems uniquely exist  \citep{bigot_2018_CharacterizationBarycentersWasserstein}. Therefore, a sufficiently small $\delta$ can be set to bound the discrepancy between $U(\nu_1)$ and $U(\nu_2)$ by any $\epsilon > 0$ as two continuous, strictly convex functionals converge as $\delta \rightarrow 0$. 
	
	Next, we show that the updating scheme induces a non-increasing sequence regarding the cost function $F$, which is equivalent to show that the inequality $F(U(\nu)) \leq F(\nu)$ holds. At iteration $t$, the IRLS minimization problem is defined as 
	\begin{equation*}
	\min_\nu G_t (\nu) = \min_\nu \sum_{n=1}^N \frac{\pi}{\calW_2 (\nu^{(t)}, \mu_n)} \calW_2^2 (\nu, \mu_n) = \min_\nu \sum_{n=1}^N \frac{\pi}{\calW_2 (\nu^{(t)}, \mu_n)} d^2 (\log_{\nu^{(t)}} (\nu), \log_{\nu^{(t)}} (\mu_n)),
	\end{equation*}
	where $d : \text{Tan}_{\nu^{(t)}} \times \text{Tan}_{\nu^{(t)}} \rightarrow \bbR$ is a distance function on the tangent space at $\nu^{(t)}$. An iterate $\nu^{(t+1)}$ is defined a minimizer of $G_t (\nu)$ so that $G_t (\nu^{(t+1)}) \leq G_t (\nu^{(t)})$ and the equality holds if $\nu^{(t)} = \nu^{(t+1)}$ due to the uniqueness of barycenter. Since $\calP_2(\bbR^d)$ has nonnegative sectional curvature and is geodesically convex, we have $	\calW_2 (\nu^{(t+1)}, \mu_n) \leq d ( \log_{\nu^{(t)}} (\nu^{(t+1)}), \log_{\nu^{(t)}} (\mu_n))$ as a consequence of the Topogonov's theorem. This leads to the following relationship, 
	\begin{eqnarray*}
		F (\nu^{(t)})  &=& \sum_{n=1}^N \pi_n \calW_2 (\nu^{(t)}, \mu_n) \\
		&\geq& \sum_{n=1}^N \frac{\pi_n}{\calW_2 (\nu^{(t)}, \mu_n)} \calW_2^2 (\nu^{(t+1)}, \mu_n)\\
		&=& \sum_{N=1}^N \frac{\left(\pi_n\calW_2 (\nu^{(t+1)}, \mu_n)\right)^2 }{\pi_n \calW_2 (\nu^{(t)}, \mu_n)},
	\end{eqnarray*}
	which can be simplified as $\sum_{n=1}^N \alpha_n \geq \sum_{n=1}^N \beta_n^2 / \alpha_n$ for $\alpha_n = \pi_n \calW_2 (\nu^{(t)}, \mu_n)$ and $\beta_n = \pi_n \calW_2 (\nu^{(t+1)}, \mu_n)$. If we define a univariate function $h(x) = \sum_{n=1}^N \alpha_n^{1-x} \beta_n^x = \sum_{n=1}^N \alpha_n (\beta_n / \alpha_n)^x$, the above relationship is paraphrased as $h(0) \geq h(2)$. Since $h$ is a convex function as $d^2 h / dx^2 \geq 0$, the following holds
	\begin{equation*}
	h(1) = h\left(\frac{1}{2}\cdot 0 + \frac{1}{2}\cdot 2\right) \leq \frac{1}{2} h(0) + \frac{1}{2} h(2) \leq h(0),
	\end{equation*}
	which proves our claim since
	\begin{equation*}
	\sum_{n=1}^N \pi_n \calW_2 (\nu^{(t+1)}, \mu_n) = h(1) \leq h(0) = \sum_{n=1}^N \pi_n \calW_2(\nu^{(t)}, \mu_n),
	\end{equation*}
	and equality holds if $\calW_2(\nu^{(t)}, \mu_n) = \calW_2(\nu^{(t+1)}, \mu_n)$ for all $n$. 
	
	Now we prove the main part of the theorem. Let $\lbrace \nu^{(t)}\rbrace$ be a sequence of updates starting from $\nu^{(0)}$. Since $F(\nu^{(0)}) < \infty$  and the IRLS updating rule engenders a non-increasing sequence, we can take a minimizing sequence $\lbrace \nu^{(t_k)}\rbrace_{k=1}^\infty$ that converges to some $\nu^*$ in the sense that $\lim_{k\rightarrow \infty}\,F(\nu^{(t_k)}) = F(\nu^*)$, which leads to the following observation that
	\begin{equation*}
	\lim_{k\rightarrow \infty}\,F(U(\nu^{(t_k)})) = F(U(\nu^*)),
	\end{equation*}
	by the continuity of an updating map. Moreover, $\lbrace F(\nu^{(t)})\rbrace_{t=1}^{\infty}$ is a bounded and non-increasing sequence in $\bbR$ so that there exists an accumulation point and it must correspond to that of a convergent subsequence. Therefore, we end up with 
	\begin{equation*}
	F(U(\nu^*)) = \lim_{k\rightarrow \infty} F(U(\nu^{(t_k)})) 	= \lim_{k\rightarrow \infty} F(\nu^{(t_k)}) = \lim_{t\rightarrow\infty} F(\nu^{(t)}) = F(\nu^*),
	\end{equation*}
	where an accumulation point $\nu^*$ of the sequence belongs to a set of stationary points $S$ as stated in the theorem.
\end{proof}

\subsubsection*{Proposition 5}
\begin{proof}
Computation of the Wasserstein median, as described in Algorithm 1, minimize a sequence of functionals, 
\begin{equation*}
	G_t (\nu) = \sum_{n=1}^N \tilde{w}_n^{(t)} \calW_2^2(\nu,\mu_n),
\end{equation*}
where this subproblem corresponds to a weighted barycenter problem. Since any weighted barycenter problem is equivalent to solving a barycenter problem where observations are replicated according to their relative weights, the Algorithm 1 engenders a unique convergent sequence $\lbrace \nu^{(t)} \rbrace $ as $t \rightarrow \infty$. Since the convergent sequence is unique, its stationary point is unique hence the attained Wasserstein median is unique as well. 
\end{proof}

\subsection*{B. Additional Simulated Examples}

Here we present extended version of the simulated examples from our manuscript onto different classes of probability measures. To rephrase, this simulated experiment aims at quantifying empirical robustness of the Wasserstein median when given a collection of probability measures. For all cases, we consider a collection of 100 probability measures, where the majority of them comes from \textit{signal} (type 1) and all the others are considered as \textit{contamination} (type 2). Each probability measure is constructed empirically by first drawing a random sample of varying sizes and then estimated accordingly. 

\subsubsection*{B.1. Centered Gaussian Measures}

\begin{figure}[ht]
	\centering
	\begin{subfigure}[b]{0.3\textwidth}
		\centering
		\includegraphics[width=\textwidth]{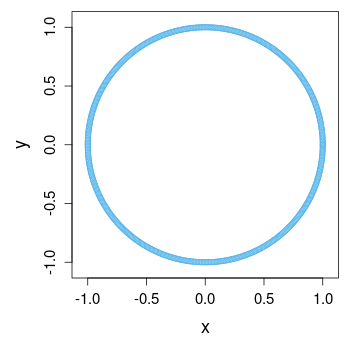}
		%\label{fig:y equals x}
	\end{subfigure}
	\hspace{0.2cm}
	\begin{subfigure}[b]{0.3\textwidth}
		\centering
		\includegraphics[width=\textwidth]{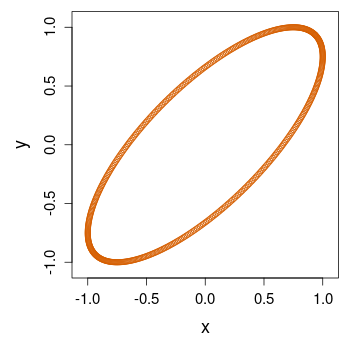}
		%\label{fig:three sin x}
	\end{subfigure}
	\caption{Two model Gaussian measures  $N({\bf \Sigma}_1)$ (left in {\color{wass-class1}  \bf light blue}) and $N({\bf \Sigma}_2)$ (right in {\color{wass-class2} \bf orange}) visualized as 95\% confidence ellipses of their covariance matrices, respectively.}
	\label{fig:gauss_twosigs}
\end{figure}

The first additional example is the case  where objects of interests are centered Gaussian distributions in $\mathbb{R}^2$. We take two Gaussian distributions $N({\bf  \Sigma}_1)$ and $N({\bf \Sigma}_2)$ as sources of \textit{signal}  and \textit{contamination} respectively. The two covariance matrices are given as 
\begin{equation*}
	{\bf \Sigma}_1 = \begin{pmatrix}
		1 & 0 \\ 0 & 1
	\end{pmatrix}\qquad\text{and} \qquad{\bf \Sigma}_2 = \begin{pmatrix}
		1 & 0.75 \\ 0.75 & 1
	\end{pmatrix},
\end{equation*}
which are graphically presented in Figure \ref{fig:gauss_twosigs}.

As shown above, an identity matrix ${\bf \Sigma}_1$ is drawn as a circle of radius 1 and ${\bf \Sigma}_2$ is a rotated ellipse. Similar to the univariate distribution example from the manuscript,  we generate perturbed variants of each distribution by computing a sample covariance for a randomly generated sample from the distribution. We repeat this process $(100-k)$ times for $N({\bf \Sigma}_1)$ and  $k$ times for $N({\bf \Sigma}_2)$ for $k\in[1,25]$. This simulates a scenario where a majority of centered Gaussian measures resembles the signal measure $N({\bf \Sigma}_1)$ and a small portion of perturbation comes from the contamination $N({\bf \Sigma}_2)$. When barycenter and median estimates are obtained, we report the discrepancy between the estimates and the signal measure $N({\bf \Sigma}_1)$ by the 2-Wasserstein distance. 

\begin{figure}[h]
	\centering
	\begin{subfigure}[b]{0.22\textwidth}
		\centering
		\caption{}
		\includegraphics[width=\textwidth]{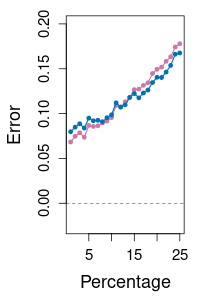}
		%\label{fig:y equals x}
	\end{subfigure}
	\hfill
	\begin{subfigure}[b]{0.22\textwidth}
		\centering
		\caption{}
		\includegraphics[width=\textwidth]{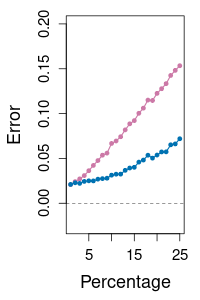}
		%\caption{}
		%\label{fig:three sin x}
	\end{subfigure}
	\hfill
	\begin{subfigure}[b]{0.22\textwidth}
		\centering
		\caption{}
		\includegraphics[width=\textwidth]{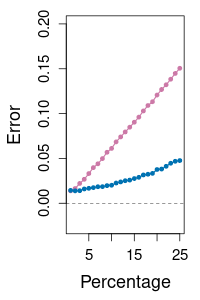}
		%\caption{}
		%\label{fig:three sin x}
	\end{subfigure}
	\hfill
	\begin{subfigure}[b]{0.22\textwidth}
		\centering
		\caption{}
		\includegraphics[width=\textwidth]{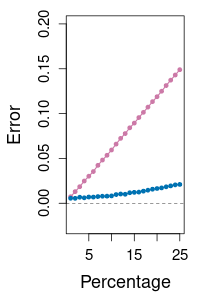}
		%\caption{}
		%\label{fig:three sin x}
	\end{subfigure}
	\caption{Performance comparison for the centered Gaussian distribution example. Average error of 50 runs is measured between the signal measure $N(\Sigma_1)$ and two centroid estimates, the Wasserstein barycenter (in {\color{wass-mean}\bf purple}) and the Wasserstein median (in {\color{wass-meds}\bf blue}), across varying degrees of contamination where the size of a random sample for covariance estimation is (a) 10, (b) 50, (c) 100, and (d) 500.}
	\label{fig:gauss_performance}
\end{figure}

\begin{figure}[h]
	\centering
	\begin{subfigure}[b]{0.16\textwidth}
		\centering
		\caption{}
		\includegraphics[width=\textwidth]{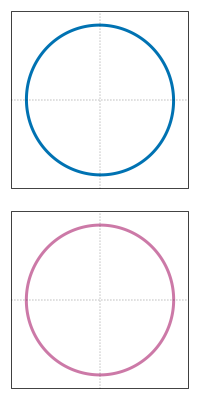}
		%\label{fig:y equals x}
	\end{subfigure}
	\begin{subfigure}[b]{0.16\textwidth}
		\centering
		\caption{}
		\includegraphics[width=\textwidth]{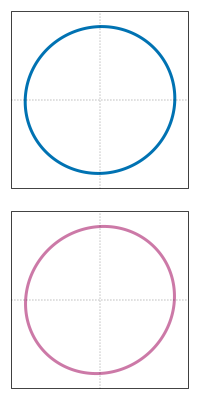}
		%\caption{}
		%\label{fig:three sin x}
	\end{subfigure}
	\begin{subfigure}[b]{0.16\textwidth}
		\centering
		\caption{}
		\includegraphics[width=\textwidth]{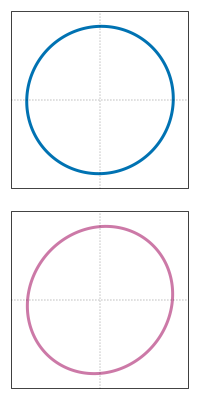}
		%\caption{}
		%\label{fig:three sin x}
	\end{subfigure}
	\begin{subfigure}[b]{0.16\textwidth}
		\centering
		\caption{}
		\includegraphics[width=\textwidth]{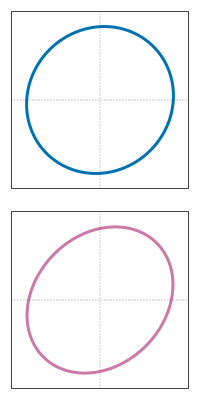}
		%\caption{}
		%\label{fig:three sin x}
	\end{subfigure}
	\caption{Visualization of the Wasserstein median (top row in {\color{wass-meds} \bf blue}) and the Wasserstein barycenter (bottom row in {\color{wass-mean} \bf purple}) given a collection of 100 centered Gaussian distributions consisting of two types. Each covariance matrix is estimated from a corresponding random sample of size 500. The degrees of contamination are (a) $1\%$, (b) $5\%$, (c) $10\%$, and (d) $25\%$.}
	\label{fig:gauss_estimates}
\end{figure}

The results are summarized in Figure \ref{fig:gauss_performance}. Except for the case where a random sample is of size 10, we see a consistent pattern as before that difference between the Wasserstein median and the signal measure remains almost constant while that of the barycenter magnifies significantly as the degree of contamination increases. This is also validated in visual inspection of covariance estimates represented as ellipses in Figure \ref{fig:gauss_estimates} that the barycenter becomes more rotated with increasing eccentricity parallel to the degree of contamination. On the other hand, the Wasserstein median remains close to the signal measure $N({\bf \Sigma}_1)$ and tends to be less deformed as shown in Figure \ref{fig:gauss_estimates}. This provides certain degree of evidence for robustness of the Wasserstein median in the presence of outliers.

\subsubsection*{B.2. High-dimensional Gaussian Measures}

\begin{figure}[ht]
	\centering
	\includegraphics[width=.8\linewidth]{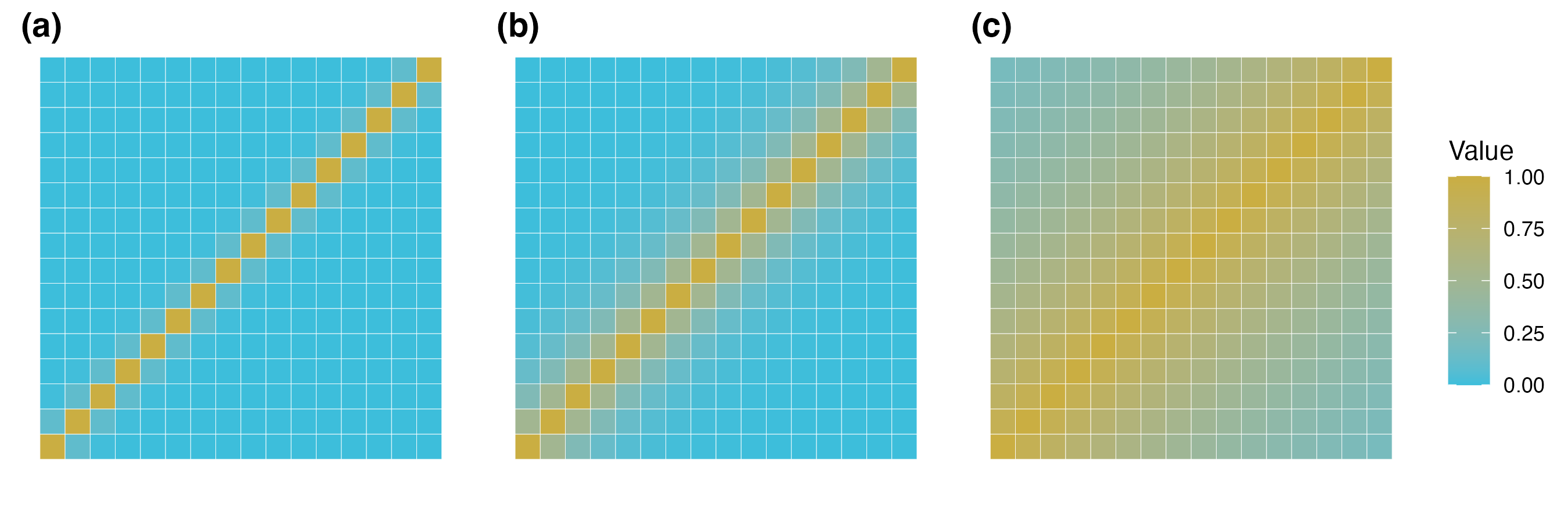}
	\caption{Visualization of three $16\times 16$ covariance matrices according to the AR(1) process for (a) $\rho=0.1$, (b) $\rho=0.5$, and (c) $\rho=0.9$.  }
	\label{fig:high3covs}
\end{figure}

We extend the centered Gaussian measures case to the high-dimensional setting. Similarly, we take two Gaussian distributions $N({\bf \Sigma_1})$ and $N({\bf \Sigma_2})$ as sources of \textit{signal} and \textit{contamination}, respectively. We consider multiple dimensions of $p=16,32,64,128,256,512$, and $1024$. Across all dimensions, we take an identity matrix as the signal covariance matrix, i.e., ${\bf \Sigma_1} = I_{p\times p}$. For the contamination, we adopt an autoregressive covariance structure as described in \cite{bickel_2008_RegularizedEstimationLarge}. That is, ${\bf \Sigma_2}$ is the covariance matrix of an AR(1) process,
\begin{equation*}
	{\bf \Sigma_2} (i,j) = \rho^{|i-j|},\quad 1\leq i,j \leq p,
\end{equation*}
for a decay parameter $\rho$ set to $0.1$, $0.5$, or $0.9$.  As shown in Figure \ref{fig:high3covs}, contamination covariance according to AR(1) with $\rho=0.1$ has off-diagonal terms that quickly decay so that the covariance matrix is almost equivalent to the identity matrix. On the other hand, larger values of $\rho$ result in off-diagonal entries being filled with values greater than 0. We selected AR(1) process covariance structures with varying $\rho$ values to investigate the differential impact of the Wasserstein centroids contingent on the degree of discrepancy between the signal and the contamination.

As before, we generated a random sample of size $2p$ from $N({\bf \Sigma_1})$ and estimated its empirical covariance, resulting in a perturbed realization of ${\bf \Sigma_1}$. This sampling and estimation process was repeated $100-k$ times. Similarly, we repeated this process $k$ times for ${\bf \Sigma_2}$. Aggregating these two sets of covariances yielded a collection of 100 covariance matrices, with $k$ representing contaminations. Then, the Wasserstein barycenter and median were computed according to the Bures-Wasserstein geometry of centered Gaussian distributions, which we denote as $\hat{\Sigma}_{\text{mean}}$ and $\hat{\Sigma}_{\text{med}}$, respectively. Our experiment considers different degrees of contamination for $k=1,5,10,25$ across varying dimensions $p=16,32,\ldots,1024$. Also, we added another dimension of comparison for $\rho=0.1, 0.5, 0.9$. For each setting, we quantified the comparative performance using the following metric
\begin{equation}\label{eq:metric}
\tau = 	\frac{\calW_2(I_{p\times p},\hat{\Sigma}_{\text{mean}} )}{\calW_2(I_{p\times p},\hat{\Sigma}_{\text{med}} )},
\end{equation}
which measures the relative accuracy of a barycenter estimate to that of a median estimate. By its construction, this implies that the higher $\tau$ is, the closer the estimated Wasserstein median is to the true signal. Results of our experiment are summarized in Table \ref{tab:HighGauss}.

\begin{table}[h]
	\centering
	\begin{subtable}{\linewidth}
		\centering
		\caption{$\rho=0.1$}
			\begin{tabular}{|c|c|c|c|c|c|c|c|}
			\hline
			\multirow{2}{*}{\begin{tabular}[c]{@{}c@{}}Noise\\ Level\end{tabular}} & \multicolumn{7}{c|}{Dimensionality}                                                                                                                                 \\ \cline{2-8} 
			& 16 & 32 & 64 & 128 & 256 & 512 & 1024 \\   \hline
			1\% & 0.992361 & 0.998510 & 0.999838 & 0.999934 & 0.999981 & 1.000002 & 1.000004 \\   5\% & 0.998741 & 0.998972 & 0.999516 & 0.999944 & 1.000019 & 1.000039 & 1.000044 \\   10\% & 0.999601 & 0.999319 & 0.999755 & 0.999934 & 1.000121 & 1.000107 & 1.000116 \\   25\% & 0.997386 & 0.999669 & 0.999623 & 0.999987 & 1.000245 & 1.000240 & 1.000234 \\ 
			\hline 
		\end{tabular}
	\end{subtable}%
	
	\begin{subtable}{\linewidth}
		\centering
		\caption{$\rho=0.5$}
			\begin{tabular}{|c|c|c|c|c|c|c|c|}
			\hline
			\multirow{2}{*}{\begin{tabular}[c]{@{}c@{}}Noise\\ Level\end{tabular}} & \multicolumn{7}{c|}{Dimensionality}                                                                                                                                 \\ \cline{2-8} 
			& 16 & 32 & 64 & 128 & 256 & 512 & 1024 \\   \hline
			1\% & 0.993565 & 1.000575 & 1.001828 & 1.002179 & 1.002243 & 1.002239 & 1.002237 \\   5\% & 1.004461 & 1.012590 & 1.018531 & 1.020313 & 1.020090 & 1.020131 & 1.020212 \\   10\% & 1.028025 & 1.039723 & 1.045105 & 1.047711 & 1.049302 & 1.049529 & 1.050064 \\   25\% & 1.054711 & 1.066210 & 1.060924 & 1.066897 & 1.069378 & 1.070116 & 1.069653 \\
			\hline 
		\end{tabular}
	\end{subtable}% 
	
	\begin{subtable}{\linewidth}
		\centering
		\caption{$\rho=0.9$}
		\begin{tabular}{|c|c|c|c|c|c|c|c|}
			\hline
			\multirow{2}{*}{\begin{tabular}[c]{@{}c@{}}Noise\\ Level\end{tabular}} & \multicolumn{7}{c|}{Dimensionality}                                                                                                                                 \\ \cline{2-8} 
			& 16 & 32 & 64 & 128 & 256 & 512 & 1024 \\   \hline
			1\% & 1.011344 & 1.020349 & 1.024466 & 1.023667 & 1.024179 & 1.024455 & 1.024649 \\   5\% & 1.138970 & 1.162608 & 1.158127 & 1.168165 & 1.169119 & 1.172395 & 1.172888 \\   10\% & 1.289637 & 1.319389 & 1.342145 & 1.344138 & 1.357256 & 1.360231 & 1.359940 \\   25\% & 1.403383 & 1.424628 & 1.473259 & 1.480404 & 1.481012 & 1.485441 & 1.488401 \\
			\hline 
		\end{tabular}
	\end{subtable}% 
	\caption{Comparison of the Wasserstein barycenter and median in the high-dimensional Gaussian measures example. Relative performance as defined in Equation \ref{eq:metric} is reported for a range of dimensions and degrees of contamination. Covariance matrices of contamination class from AR(1) process are separately tested for (a) $\rho=0.1$, (b) $\rho=0.5$, and (c) $\rho=0.9$.}
	\label{tab:HighGauss}
\end{table}

When $\rho=0.1$, the Wasserstein median performed slightly worse than the barycenter for all noise levels up to $p=128$, though this trend reversed for $p > 128$. Nevertheless, the relative performance metric at $\rho=0.1$ does not deviate significantly from 1, suggesting that both estimates behave similarly.  This similarity is due to the fact that the covariance matrices from the signal and contamination are very similar, as shown in Figure \ref{fig:high3covs}. As the contamination becomes more predominant, the Wasserstein median quickly gains supremacy over the barycenter. In the case of $\rho=0.5$, the Wasserstein median outperformed the barycenter in all but one scenario, where only one contamination measure exists. The margin becomes even more pronounced in the $\rho=0.9$ case, with the gain of the Wasserstein median reaching up to 48\%. We also observed a universal pattern: the Wasserstein median demonstrates closer proximity to the signal measure as the degree of contamination increases - namely, as $k$ becomes larger across all $\rho$ values and dimensions. This provides empirical evidence supporting our claim that the Wasserstein median is indeed a robust alternative to the Wasserstein barycenter.

\subsubsection*{B.3. Histograms}

Histogram is another popular representation of the distribution, which can be considered as a discrete measure supported on one-dimensional lattice by considering the midpoints of each bin as points of the lattice. Similar to before, we consider a signal-noise setting to see applicability of the proposed framework on a collection of histogram objects. In this example, we opt for the beta distribution denoted as Beta($\alpha,\beta$) for two shape parameters $\alpha > 0$ and $\beta > 0$ as a model distribution. We consider Beta(2,5) as the signal measure whose density function has a mound shape and positive skewness. For the contamination measure, we use Beta(5,1) whose density function is monotonically increasing in the bounded support. We use an equidistant partition $0=t_0 < \cdots < t_{20} = 1$ for binning a randomly generated sample and normalize bin counts as relative frequency so that the binned vector sums to 1.

\begin{figure}[ht]
	\centering
	\begin{subfigure}[b]{0.17\textwidth}
		\centering
		\caption{}
		\includegraphics[width=\textwidth]{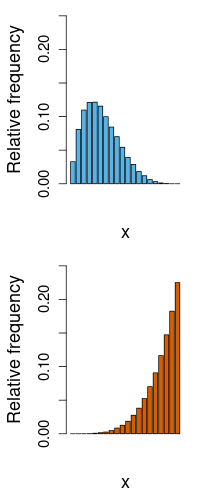}
		%\label{fig:y equals x}
	\end{subfigure}
	\begin{subfigure}[b]{0.17\textwidth}
		\centering
		\caption{}
		\includegraphics[width=\textwidth]{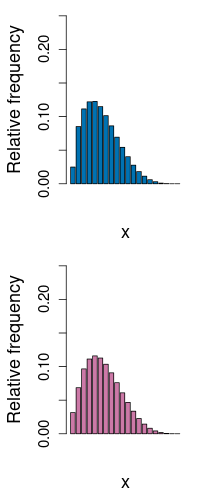}
		%\label{fig:y equals x}
	\end{subfigure}
	\begin{subfigure}[b]{0.17\textwidth}
		\centering
		\caption{}
		\includegraphics[width=\textwidth]{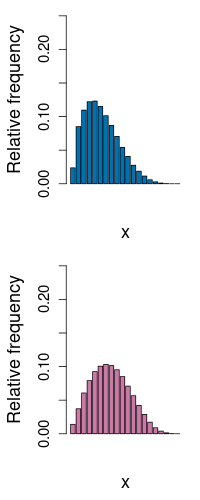}
		%\caption{}
		%\label{fig:three sin x}
	\end{subfigure}
	\begin{subfigure}[b]{0.17\textwidth}
		\centering
		\caption{}
		\includegraphics[width=\textwidth]{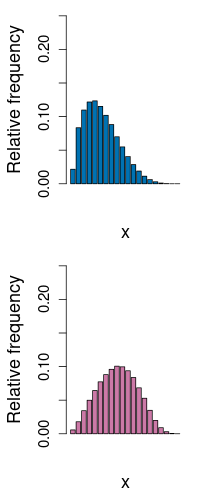}
		%\caption{}
		%\label{fig:three sin x}
	\end{subfigure}
	\begin{subfigure}[b]{0.17\textwidth}
		\centering
		\caption{}
		\includegraphics[width=\textwidth]{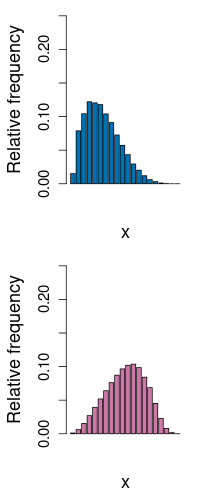}
		%\caption{}
		%\label{fig:three sin x}
	\end{subfigure}
	\caption{Visualization of the model beta distributions and the centroid estimates. The first column (a) presents density functions for the signal  Beta(2,5) (top in {\color{wass-class1} \bf light blue})	and the contamination Beta(5,1) (bottom in {\color{wass-class2} \bf orange}). For the rest, estimated histograms of Wasserstein median (top row in {\color{wass-meds} \bf blue}) and the Wasserstein barycenter (bottom row in {\color{wass-mean} \bf purple}) are shown where each column corresponds to the varying degrees of contamination at (b) $1\%$, (c) $5\%$, (d) $10\%$, and (e) $25\%$. Each random sample has the size of 250. }
	\label{fig:histogram_estimates}
\end{figure}

\begin{figure}[h]
	\centering
	\begin{subfigure}[b]{0.20\textwidth}
		\centering
		\caption{}
		\includegraphics[width=\textwidth]{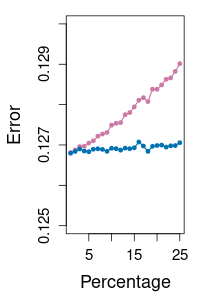}
		%\label{fig:y equals x}
	\end{subfigure}
	\hfill
	\begin{subfigure}[b]{0.20\textwidth}
		\centering
		\caption{}
		\includegraphics[width=\textwidth]{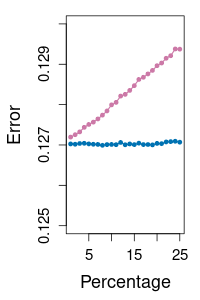}
		%\caption{}
		%\label{fig:three sin x}
	\end{subfigure}
	\hfill
	\begin{subfigure}[b]{0.20\textwidth}
		\centering
		\caption{}
		\includegraphics[width=\textwidth]{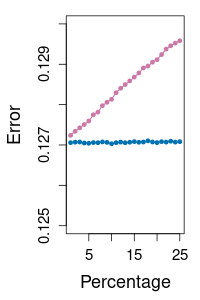}
		%\caption{}
		%\label{fig:three sin x}
	\end{subfigure}
	\hfill
	\begin{subfigure}[b]{0.20\textwidth}
		\centering
		\caption{}
		\includegraphics[width=\textwidth]{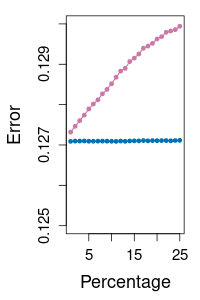}
		%\caption{}
		%\label{fig:three sin x}
	\end{subfigure}
	\caption{Performance comparison for the histogram example. Average error of 50 runs is measured between the signal measure Beta(2,5) and two centroid estimates, the Wasserstein median (in {\color{wass-meds}\bf blue}) and the Wasserstein barycenter (in {\color{wass-mean} \bf purple}), across varying degrees of contamination where a random sample is of size (a) 10, (b) 50, (c) 100, and (d) 500.}
	\label{fig:histogram_performance}
\end{figure}

The designated model measures are shown in Figure \ref{fig:histogram_estimates} along with two centroid estimates across multiple levels of contamination. As the degree of contamination increases, estimated barycenters tend to deviate more from the signal measure while the medians remain nearly identical. This can be viewed in the sense of skewness where the barycenter becomes more negatively skewed for higher levels of  contamination. Hence, we may argue that the barycenter altogether fails to characterize one of the basic properties for a desired measure of central tendency given a set of histograms. Our extended experiment to have 1 to 25 contaminants among a total of 100 histograms leads to the similar observation as presented in Figure \ref{fig:histogram_performance} that the median outperforms the barycenter across all settings.

\bibliographystyle{dcu}
\bibliography{main}

\end{document}